\documentclass[showpacs,aps,prb,floatfix,twocolumn,superscriptaddress]{revtex4}

\usepackage{graphicx}
\usepackage{dcolumn}
\usepackage{bm}
\usepackage{amsmath}

\usepackage{array}
\usepackage{color}
\usepackage{float}
\usepackage{subfigure}
\usepackage{dsfont}
\usepackage{txfonts}
\usepackage{wasysym}
\usepackage{multirow}

\newcommand{\bk}{\mathbf{k}}

\newcommand{\dn}{\downarrow}

\newcommand{\e}{\varepsilon}
\newcommand{\el}{\mathrm{e}}
\newcommand{\half}{{\textstyle\frac{1}{2}}}
\newcommand{\Hol}{\mathrm{Hol}}
\newcommand{\leads}{\mathrm{leads}}
\newcommand{\mol}{\mathrm{mol}}
\newcommand{\molleads}{\mathrm{mol-leads}}
\newcommand{\ph}{\mathrm{ph}}
\newcommand{\pdag}{\phantom{\dag}}
\newcommand{\s}{\sigma}
\newcommand{\sbar}{\bar{\s}}
\newcommand{\tot}{\mathrm{tot}}
\newcommand{\tun}{\mathrm{tun}}
\newcommand{\up}{\uparrow}

\begin{document}
\title{Effects of electron-phonon coupling in the Kondo regime of a
two-orbital molecule}

\author{G.\ I.\ Luiz}
\affiliation{Instituto de F\'isica, Universidade Federal de Uberl\^andia,
Uberl\^andia, MG 38400-902, Brazil}
\author{E.\ Vernek}
\affiliation{Instituto de F\'isica, Universidade Federal de Uberl\^andia,
Uberl\^andia, MG 38400-902, Brazil}
\author{L.\ Deng}
\affiliation{Department of Physics, University of Florida, P.O.\ Box 118440,
Gainesville, Florida 32611-8440, USA}
\author{K.\ Ingersent}
\affiliation{Department of Physics, University of Florida, P.O.\ Box 118440,
Gainesville, Florida 32611-8440, USA}
\author{E.\ V.\ Anda}
\affiliation{Departamento de F\'{\i}sica, Pontif\'{\i}cia Universidade
Cat\'olica do Rio de Janeiro, RJ 22453-900, Brazil}
\date{\today}
\begin{abstract}
\end{abstract}



\begin{abstract}
We study the interplay between strong electron-electron and electron-phonon
interactions within a two-orbital molecule coupled to metallic leads, taking
into account Holstein-like coupling of a local phonon mode to the molecular
charge as well as phonon-mediated interorbital tunneling. By combining canonical
transformations with numerical renormalization-group calculations to address
the interactions nonperturbatively and on equal footing, we obtain a
comprehensive description of the system's many-body physics in the
anti-adiabatic regime where the phonons adjust rapidly to changes in the
orbital occupancies, and are thereby able to strongly affect the Kondo physics.
The electron-phonon interactions strongly modify the bare orbital energies and
the Coulomb repulsion between electrons in the molecule, and tend to inhibit
tunneling of electrons between the molecule and the leads. The consequences of
these effects are considerably more pronounced when both molecular orbitals lie
near the Fermi energy of the leads than when only one orbital is active.
In situations where a local moment forms on the molecule, there is a crossover
with increasing electron-phonon coupling from a regime of collective Kondo
screening of the moment to a limit of local phonon quenching. At low
temperatures, this crossover is associated with a rapid increase in the
electronic occupancy of the molecule as well as a marked drop in the linear
electrical conductance through the single-molecule junction.
\end{abstract}

\date{\today}
\maketitle

\section{Introduction}

Single-molecule junctions\cite{NanoLett.6.1784,PhysRevB.72.033408,
J.Chem.Phys..117.11033,J.Chem.Phys..128.154706} are structures consisting of a
single molecule bridging the gap between source and drain electrodes, allowing
electronic transport when a bias voltage is applied across the structure.
These systems, which manifest a rich variety of experimentally accessible
physics in a relatively simple setting,\cite{NatureCommunications.2.305} have
attracted much theoretical and experimental effort in connection with molecular
electronics.\cite{Science.272.698,Phys.Today.56.43} A major goal of these
efforts has been to take advantage of natural or artificial molecules for
technological purposes. Examples of single-molecule junctions encompass, for
example, single hydrogen molecules\cite{Nature.419.906,PhysRevB.71.161402,
PhysRevB.77.115326} and more complex structures such as $4,4'$-bipyridine
molecules coupled to metallic nanocontacts.\cite{PhysRevB.78.165116,
PhysRevB.73.075326,PhysRevB.72.241401}

An important ingredient in transport through molecular systems is the
electron-electron interaction (Coulomb repulsion), the effect of which is
greatly enhanced by the spatial confinement of electrons in molecules.
Electron-electron (e-e) interactions are known to produce Coulomb
blockade phenomena\cite{Tans:98,Park:02,Kubatkin:03}
and Kondo correlations\cite{Hewson-Kondo,Nygard:00,Park:02,
Liang:02,Pasupathy:04} at low temperatures.
Confined electrons are also known to couple to quantized vibrations (phonons)
of the molecules,\cite{PhysRevLett.17.1139} resulting in important effects on
electronic transport,\cite{PhysRevB.14.3177,Science.280.1732,Park:00,
PhysRevLett.85.1918,Weig:04,NanoLett.4.639} including
vibrational side-bands found at finite bias in the Kondo regime.\cite{Yu:04,
Parks:07,Fernandez-Torrente:08} Single-molecule junctions therefore provide
a valuable opportunity to study charge transfer in systems with strong competing
interactions.\cite{Galperin:07,Hartle:11}

It has recently been demonstrated that the energies of the molecular orbitals
in a single-molecule junction can be tuned relative to the Fermi energy of the
electrodes by varying the voltage applied to a capacitively coupled
gate.\cite{Nature.462.1039} Similar control has for some time been available
in another class of nanoelectronic device: a quantum dot coupled to a
two-dimensional electron gas.\cite{Nature.391.156,Science.293.2221} The
electrons confined in a quantum dot couple---in most cases quite weakly---to
collective vibrations of the dot and its substrate.\cite{PhysRevB.49.13704}
In single-molecule devices, by contrast, the confined electrons interact with
local vibration modes of the molecule that can produce pronounced changes in
the molecular electronic orbitals. Consequently, electron-phonon (e-ph)
interactions are expected to play a much more important role in molecules
than in quantum dots.

From the theoretical point of view, addressing both e-e and e-ph interactions
from first principles is a very complicated task. However, simple effective
models can provide good qualitative results, depending on the parameter regime
and the method employed to solve the model
Hamiltonian.\cite{Galperin:07} For example, the essential
physics of certain experiments\cite{Park:00,Weig:04}
appears to be described by variants of the Anderson-Holstein model, which
augments the Anderson model\cite{Anderson:61} for a magnetic impurity in a
metallic host with a Holstein coupling\cite{Holstein:59} of the impurity charge
to a local phonon mode. Variants of the model have been studied since the 1970s
in connection with other problems\cite{Simanek:79,Ting:80,Kaga:80,
Schonhammer:84,Alascio:88,Schuttler:88,Ostreich:91,Hewson:02,Jeon:03,Zhu:03,
Lee:04} prior to their application to single-molecule
devices.\cite{Cornaglia:04,Cornaglia:05,Paaske:05,Mravlje:05,
PhysRevLett.95.256807,Mravlje:06,Nunez:07,Cornaglia:07,Mravlje:08,Dias:09}
Various analytical approximations as well as nonperturbative numerical
renormalization-group calculations have shown that in equilibrium, the
Holstein coupling reduces the Coulomb repulsion between two electrons in the
impurity level, even yielding effective e-e attraction for sufficiently
strong e-ph coupling. Increasing the e-ph coupling from zero can produce a
smooth crossover from a conventional Kondo effect, involving conduction-band
screening of the impurity spin degree of freedom, to a charge Kondo effect in
which it is the impurity ``isospin'' or deviation from half-filling that is
quenched by the conduction band. In certain cases, the system may realize
the two-channel Kondo effect.\cite{Dias:09}

Single-molecule devices at finite bias are usually studied via nonequilibrium
Keldysh Green's functions that systematically incorporate the many-body
interactions within a system. Although this approach has proved to be the most
reliable for calculation of transport properties, the results are highly
sensitive to the approximations made. For instance, the equation-of-motion
technique\cite{PhysRevB.82.235412} generates a hierarchy of coupled equations
for Green's functions containing $2n$ fermionic operators for $n=1$, $2$, $3$,
$\ldots$: a hierarchy that must be decoupled at some level in order to become
useful. The most commonly used decoupling scheme is based on a mean-field
decomposition of the $n=2$ Green's functions, leading to the well-known Hubbard
I approximation.\cite{Proc.Roy.Soc.(London).A276.238} This approximation give
reasonable results for temperatures $T$ above the system's Kondo temperature
$T_K$, but as it neglects spin correlations between localized and conduction
electrons, it fails in the Kondo regime.

A few years ago, two of us applied the equation-of-motion method decoupled at
level $n=2$ to study a single-molecule junction that features phonon-assisted
interorbital tunneling.\cite{PhysRevB.76.075320} However, to capture the
physics at $T\lesssim T_K$ requires extension of the equation-of-motion
hierarchy to higher order, which in most cases is carried out in the limit of
infinite Coulomb interaction. The Kondo regime may also be studied
via diagrammatic expansion within the non-crossing approximation, which is
again most straight forward in the infinite-interaction
limit.\cite{J.Phy.:Condens.Matter.23.125302}

This paper reports the results of an investigation of the Kondo regime of a
two-orbital molecule, with focus on situations in which Coulomb interactions
are strong but finite. Our model Hamiltonian, which includes both
phonon-assisted interorbital tunneling and a Holstein-type coupling between the
molecular charge and the displacement of the local phonon mode, may also be used
to describe two-level quantum dots or a coupled pair of single-level dots. In
order to treat e-e and e-ph interactions on an equal basis, we employ Wilson's
numerical renormalization-group approach,\cite{RevModPhys.47.773,KWW:1980,
RevModPhys.80.395} which provides complete access to the equilibrium behavior
and linear response of the system for temperatures all the way to absolute zero.
We show that the renormalization of e-e interactions is strongly dependent on
the energy difference between the two molecular orbitals.
For small interorbital energy differences, the renormalization is significantly
enhanced compared with the situation of one active molecular orbital considered
in previous work. This enhancement is detrimental for realization of the Kondo
effect but improves the prospects for accessing a phonon-dominated regime of
effective e-e attraction. A sharp crossover between Kondo and phonon-dominated
regimes, which has its origin in a level crossing that occurs when the molecule
is isolated from the leads, has signatures in thermodynamic properties and in
charge transport through the system.

Understanding the equilibrium and linear-response properties of this model is
an important precursor to studies of the nonequilibrium steady state, where
e-ph effects are likely to reveal themselves at finite bias.\cite{Yu:04,Parks:07,
Fernandez-Torrente:08} Moreover, the model we address can readily be adapted to
treat the coupling of a single-molecule junction to electromagnetic radiation,
a situation where driven interorbital transitions is likely to be of particular
importance.

The rest of the paper is organized as follows: Section \ref{sec:model}
describes our model system and provides a preliminary analysis via
canonical transformations. Section \ref{sec:method} reviews the numerical
solution method and Sec.\ \ref{sec:results} presents and analyzes results for
cases of large and small energy differences between the two molecular
orbitals. The main results are summarized in Sec.\ \ref{sec:summary}.

\section{Model and Preliminary Analysis}
\label{sec:model}

\subsection{Model Hamiltonian}
\label{subsec:model}

\begin{figure}[tb]
\centering\includegraphics[width=3.3in]{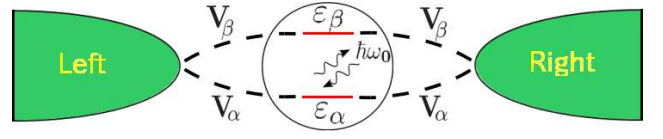}
\caption{\label{setup} (Color online)
Schematic representation of the model studied in this work. A molecule
with two active orbitals ($\alpha$ and $\beta$) spans the gap between left and
right electrodes. The molecular orbitals are subject both to e-e
and e-ph interactions.}
\end{figure}

We consider a system composed of a two-orbital molecule interacting with a local
phonon mode and also coupled to two metallic leads, as shown schematically in
Fig.~\ref{setup}. This system is modeled by the Anderson-type Hamiltonian
\begin{equation}
\label{H}
H=H_{\mol}+H_{\leads}+H_{\molleads},
\end{equation}
with $H_{\mol}$ describing the isolated molecule, $H_{\leads}$
modeling the leads, and $H_{\molleads}$ accounting for electron tunneling
between the molecule and the leads.

The molecular Hamiltonian can in turn be divided into four parts:
$H_{\mol}=H_{\el}+H_{\ph}+H_{\Hol}+H_{\tun}$. Here, the electronic part
\begin{equation}
\label{H_el}
H_{\el}=\sum_{i=\alpha,\beta} \Bigl( \e_i n_i + U_i n_{i\up}
 n_{i\dn} \Bigr) + U' n_{\alpha} n_{\beta},
\end{equation}
where $n_{i\s}=d_{i\s}^{\dag}d_{i\s}^{\pdag}$ is the number operator for
electrons of energy $\e_i$ and spin $\s$ in molecular orbital $i=\alpha$ or
$\beta$, $n_i = n_{i\up}+n_{i\dn}$, and $U_i$ and $U'$ parametrize
intraorbital and interorbital Coulomb repulsion, respectively. Without loss
of generality, we take $\e_{\beta} \ge \e_{\alpha}$. The phonon part
\begin{equation}
\label{H_ph}
H_{\ph} = \hbar\omega_0 \: n_b
\end{equation}
describes a dispersionless optical phonon mode of energy $\hbar\omega_0$, with
$n_b = b^{\dag}b$. The remaining two parts of $H_{\mol}$ describe two different
types of e-ph interaction:
\begin{equation}
\label{H_Hol}
H_{\Hol}=\lambda n_{\mol} \bigl(b^{\dag}+b\bigr)
\end{equation}
is a Holstein coupling between the phonon displacement and the combined
occupancy (i.e., charge)
\begin{equation}
n_{\mol}=n_{\alpha}+n_{\beta}
\end{equation}
of the two molecular orbitals, while
\begin{equation}
\label{H_tun}
H_{\tun}=\lambda'\sum_{\s}\Bigl(d_{\alpha\s}^{\dag}d_{\beta\s}^{\pdag}
 +d_{\beta\s}^{\dag}d_{\alpha\s}^{\pdag}\Bigr) \bigl(b^{\dag}+b\bigr)
\end{equation}
describes interorbital tunneling accompanied by emission or absorption of a
phonon. Without loss of generality, we take $\lambda\ge 0$ (since a negative
sign can be absorbed into a redefinition of the operator $b$), but we allow
$\lambda'$ to be of either sign.

The left ($\ell=L$) and right ($\ell=R$) leads are represented by
\begin{equation}
\label{H_leads}
H_{\leads}=\sum_{\ell,\bk,\s}\e_{\bk}c_{\ell\bk\s}^{\dag}c_{\ell\bk\s}^{\pdag},
\end{equation}
where $c_{\ell\bk\s}$ annihilates an electron with energy $\e_{\bk}$,
wave vector $\bk$, and spin $\s$ in lead $\ell$.
For simplicity, each lead is characterized by a flat density of states
\begin{equation}
\label{dos}
\rho(\e)=N_s^{-1}\sum_{\bk}\delta(\e-\e_{\bk})=\frac{1}{2D} \, \Theta(D-|\e|),
\end{equation}
where $N_s$ is the number of lattice sites in each lead, $D$ is the half
bandwidth and $\Theta(x)$ is the Heaviside function.

Lastly,
\begin{equation}
\label{H_molleads}
H_{\molleads}=\frac{1}{\sqrt{N_s}}\sum_{i=\alpha,\beta} \: \sum_{\ell,\bk,\s}
 \Bigl( V_{\ell i} d_{i\s}^{\dag} c_{\ell\bk\s}^{\pdag}
 + V_{i\ell} c_{\ell\bk\s}^{\dag}d_{i\s}^{\pdag}\Bigr)
\end{equation}
describes tunneling or hybridization between the molecular orbitals and the
leads, allowing transport through the system. We assume that the tunneling matrix
elements are real and have left-right symmetry so we can write
$V_{\ell i}=V_{i\ell}=V_i$. Then it is useful to perform an even/odd
transformation
\begin{align}
c_{e\bk\s}&=\frac{1}{\sqrt{2}}\bigl(c_{R\bk\s}+c_{L\bk\s} \bigr) \\
c_{o\bk\s}&=\frac{1}{\sqrt{2}}\bigl(c_{R\bk\s}-c_{L\bk\s} \bigr),
\end{align}
which allows Eq.\ \eqref{H_molleads} to be rewritten
\begin{equation}
\label{H_molleads:mod}
H_{\molleads}=\sqrt{\frac{2}{N_s}} \sum_{i=\alpha,\beta} V_i \sum_{\bk,\s}
 \Bigl(d_{i\s}^{\dag} c_{e\bk\s}^{\pdag}
 +c_{e\bk\s}^{\dag}d^{\pdag}_{i\s}\Bigr),
\end{equation}
With this transformation, the odd-parity degrees of freedom are fully decoupled
from the molecular orbitals, and can safely be dropped. As a result, the
problem reduces to one effective conduction channel described by a modified
\begin{equation}
\label{H_leads:mod}
H_{\leads}=\sum_{\bk,\s}\e_{\bk} c_{e\bk\s}^{\dag}
 c_{e\bk\s}^{\pdag} .
\end{equation}
This channel is still described by the density of states in Eq.\ \eqref{dos},
and it imparts to molecular orbital $i$ a width
\begin{equation}
\label{Gamma_i}
\Gamma_i=\pi V_i^2 / D.
\end{equation}
A similar transformation to an effective one-channel model can be derived
in any situation where the tunneling matrix elements satisfy
$V_{L\alpha}V_{R\beta}=V_{L\beta}V_{R\alpha}$, ensuring that both molecular
orbitals couple to the same linear combination of left- and right-lead states.

The Hamiltonian \eqref{H} may also describe certain quantum-dot systems. In
this setting, the ``orbitals'' $\alpha$ and $\beta$ can be taken to describe
either two active levels within a single quantum dot or the sole active level
in two different dots that are coupled to the same pair of external leads.

Since the model laid out above contains eleven energy parameters, it is
important to consider the relative values of these parameters in real
systems.  For small molecules containing up to a few hundred atoms, the largest
energy scale (apart possibly from the half bandwidth) is the local Coulomb
interaction or charging energy, which is generally of order electron volts. In
carbon nanotubes, by contrast, the charging energy can be as low\cite{Bomze:10}
as 3--4\,meV. The numerical results presented in Sec.\ IV were obtained for the
special case of equal intraorbital Coulomb repulsions $U_{\alpha}=U_{\beta}=U$
as well as equal orbital hybridizations $V_{\alpha}=V_{\beta}=V$ (and hence
orbital broadenings $\Gamma_{\alpha}=\Gamma_{\beta}=\Gamma$). These choices
prove convenient for the algebraic analysis carried out in Secs.\
\ref{subsec:preliminary} and \ref{sec:results}, but qualitatively very similar
behavior is obtained for more general ratios $U_{\beta}/U_{\alpha}$ and
$V_{\beta}/V_{\alpha}$. Most of the numerical data were computed for an
intraorbital interaction $U_{\alpha}=U_{\beta}=U=0.5D$ with an interorbital
interaction $U'$ of similar size. However, we also include a few results for the
limiting cases $U=U'=0$ and $U=U'=5$.

In the limit where one of the molecular orbitals ($\beta$, say) is removed or
becomes decoupled from the rest of the system, the Hamiltonian \eqref{H}
reduces to the Anderson-Holstein Hamiltonian.\cite{Simanek:79,Ting:80,
Kaga:80,Schonhammer:84,Alascio:88,Schuttler:88,Ostreich:91,Hewson:02,Jeon:03,
Zhu:03,Lee:04,Cornaglia:04,Cornaglia:05} It is well-established for this model
that the ratio $\hbar\omega_0/\Gamma$ is a key quantity governing the interplay
between e-ph interactions and the Kondo effect. In the \textit{instantaneous}
or \textit{anti-adiabatic} regime $\hbar\omega_0\gg\Gamma$, the bosons adjust
rapidly to any change in the orbital occupancy, leading to polaronic shifts in
the orbital energy and in the Coulomb interaction and to exponential
suppression of certain virtual tunneling processes. In the \textit{adiabatic}
regime $\hbar\omega_0\ll\Gamma$, by contrast, the phonons are unable to adjust
on the typical time scale of hybridization events, and therefore have little
impact on the Kondo physics. We expect similar behavior in the two-orbital
single-molecule junction, and concentrate in this paper solely on the
anti-adiabatic regime $U\gg\hbar\omega_0\gg\Gamma$.

In most experiments on molecular junctions, both the phonon energy\cite{Park:00,
Heersche:06,Fernandez-Torrente:08,Franke:12} and the orbital level broadening
due to the leads\cite{Liang:02,Pasupathy:04,Yu:04,Makarovski:07,
Fernandez-Torrente:08} have been found to lie in the range 5--100\, meV. All our
numerical calculations were performed for a phonon energy $\hbar\omega_0=0.1D$
and a hybridization matrix element $V=0.075$, yielding an orbital width
$\Gamma=\pi V^2/D \simeq 0.0177$ and a ratio $\omega_0/\Gamma\simeq 6$ that is
somewhat larger than---but not out of line with---that found in one of the few
experiments\cite{Fernandez-Torrente:08} that has clearly observed vibrational
effects in the Kondo regime: transport through a single tetracyanoquinodimethane
molecule, where $\omega_0=41$\,meV and $\Gamma=11$--22\,meV. Moderate control of
both $\Gamma$ and $\omega_0$ has been demonstrated in single-molecule break
junctions by changing the separation between the two electrodes,\cite{Parks:07}
so it seems probable that anti-adiabatic regime will be accessible in future
experiments.

Two other important energy scales are the e-ph couplings $\lambda$ and
$\lambda'$ (or, as will be seen below, the corresponding orbital energy shifts
$\lambda^2/\hbar\omega_0$ and $\lambda'^2/\hbar\omega_0$).
We are aware of no direct measurements of e-ph couplings in single-molecule
devices. However, first-principles calculations for one particular setup (a 1,4
benzenedithiolate molecule between aluminum electrodes) have yielded values
corresponding in our notation to $\lambda^2/\omega_0$ up to 0.02 at zero bias
and up to 0.08 at strong bias.\cite{Sergueev:05} On this basis, we believe that
it is very reasonable to consider values of $\lambda^2/\omega_0$ and
$\lambda'^2/\omega_0$ as large as 0.1.

Also important are the orbital energies $\e_{\alpha}$ and $\e_{\beta}$. Many
experimental setups allow essentially rigid shifts of these energies through
tuning of a back-gate voltage, so we consider sweeps of this form in
Sec.\ \ref{subsec:small-delta}. The energy difference $\e_{\beta}-\e_{\alpha}$
will vary widely from system to system, but is not so readily susceptible to
experimental control.

It is impossible in a paper of this length to attempt a complete exploration of
the model's parameter space. Instead, guided by the preceding discussion of
energy scales, we focus on a few representative examples that illustrate
interesting and physically relevant regimes of behavior.

\subsection{Preliminary analysis via canonical transformation}
\label{subsec:preliminary}

Insight can be gained into the properties of the two-orbital model by
performing a canonical transformation of the Lang-Firsov
type\cite{Zh.Eksp.Teor.Fiz..43.1843} from the original
Hamiltonian \eqref{H} to $\tilde{H}=e^{S_1} H e^{-S_1}$.
Following extensive algebra, it can be shown that the choice
\begin{equation}
\label{S_1}
S_1=\frac{\lambda}{\hbar\omega_0} n_{\mol} \bigl(b^{\dag}-b\bigr)
\end{equation}
eliminates the Holstein coupling between the local phonons and the molecular
electron occupancy [Eq.\ \eqref{H_Hol}], leaving a transformed Hamiltonian
\begin{equation}
\label{tilde H}
\tilde{H}=\tilde{H}_{\el}+H_{\ph}+\tilde{H}_{\tun}+H_{\leads}
 +\tilde{H}_{\molleads},
\end{equation}
in which $H_{\ph}$ and $H_{\leads}$ remain as given in Eqs.\ \eqref{H_ph} and
\eqref{H_leads:mod}, respectively; $\tilde{H}_{\el}$ has the same form as
$H_{\el}$ [Eq.\ \eqref{H_el}] with the replacements
\begin{subequations}
\label{tilde quantities}
\begin{align}
\label{tilde e_i}
\e_i&\to\tilde{\e}_i=\e_i-{\lambda^2}/{\hbar\omega_0} , \\
\label{tilde U_i}
U_i&\to\tilde{U}_i=U_i-{2\lambda^2}/{\hbar\omega_0}, \\
\label{tilde Uprime}
U'&\to\tilde{U'}=U'-{2\lambda^2}/{\hbar\omega_0} ;
\end{align}
\end{subequations}
the interorbital tunneling maps to
\begin{equation}
\label{tilde H_tun}
\tilde{H}_{\tun}=\lambda'\sum_{\s} \bigl(d_{\alpha\s}^{\dag}
 d_{\beta\s}^{\pdag} +d_{\beta\s}^{\dag} d_{\alpha\s}^{\pdag} \bigr)
 \biggl[b^{\dag}+b-\frac{2\lambda}{\hbar\omega_0}
  \bigl(1+n_{\alpha\sbar}+n_{\beta\sbar}\bigr) \biggr]
\end{equation}
where $\sbar=-\s$; and the molecule-leads coupling term becomes
\begin{equation}
\label{tilde H_molleads}
\tilde{H}_{\molleads}=\sqrt{\frac{2}{N_s}} \sum_{i=\alpha,\beta} V_i
 \sum_{\bk,\s}\Bigl(B_{\lambda}^{\dag}d_{i\s}^{\dag} c_{e\bk\s}^{\pdag}
 +B_{\lambda}^{\pdag}c_{e\bk\s}^{\dag}d_{i\s}^{\pdag} \Bigr),
\end{equation}
with
\begin{equation}
\label{B}
B_{\xi}=\exp\biggl[-\frac{\xi}{\hbar\omega_0}
 \bigl(b^{\dag}-b\bigr)\biggr] \equiv B_{-\xi}^{\dag} .
\end{equation}

This transformation extends the one applied previously (e.g., see, Ref.\
\onlinecite{Hewson:02}) to the Anderson-Holstein model. It effectively
eliminates the Holstein Hamiltonian term [Eq.\ \eqref{H_Hol}] by mapping the
local phonon mode to a different displaced oscillator basis for each value of
the total molecular occupancy $n_{\mol}$, namely, the basis that
minimizes the ground-state energy of $H_{\el}+H_{\ph}+H_{\Hol}$. There are
three compensating changes to the Hamiltonian:
(1) Shifts in the orbital energies [Eq.\ \eqref{tilde e_i}] and, more notably,
a reduction in the magnitude---or even a change in the sign---of each
e-e interaction within the molecule [Eqs.\ \eqref{tilde U_i}
and \eqref{tilde Uprime}].
These renormalizations reflect the fact that the Holstein
coupling lowers the energy of doubly occupied molecular orbitals relative to
singly occupied and empty orbitals. This well-known effect underlies the
standard e-ph mechanism for superconductivity.
(2) Addition of correlated (molecular-occupation-dependent) interorbital
tunneling [the $\lambda$-dependent term in Eq.\ \eqref{tilde H_tun}] to the
phonon-assisted tunneling present in the original Hamiltonian.
(3) Incorporation into the molecule-leads coupling [Eq.\
\eqref{tilde H_molleads}] of operators $B_{\lambda}^{\pdag}$ and
$B_{\lambda}^{\dag}$ that cause each electron tunneling event to be accompanied
by the creation and absorption of a packet of phonons as the local bosonic mode
adjusts to the change in the total molecular occupancy $n_{\mol}$.

The effects of the phonon-assisted interorbital tunneling term $H_{\tun}$ can
be qualitatively understood by rewriting Eq.\ \eqref{tilde H} in terms of even
and odd linear combinations of the $\alpha$ and $\beta$ molecular orbitals:
\begin{equation}
\label{d_p}
 d_{e\s}=\frac{1}{\sqrt{2}}\bigl(d_{\alpha\s}+d_{\beta\s}\bigr), \quad
 d_{o\s}=\frac{1}{\sqrt{2}}\bigl(d_{\alpha\s}-d_{\beta\s}\bigr).
\end{equation}
In this parity basis, Eq.\ \eqref{tilde H_tun} becomes
\begin{equation}
\label{tilde H_tun:mod}
\tilde{H}_{\tun}=\lambda' \sum_{\s} (n_{e\s} - n_{o\s})
 \biggl[b^{\dag}+b-\frac{2\lambda}{\hbar\omega_0}
  \bigl(1+n_{e\sbar}+n_{o\sbar}\bigr) \biggr] ,
\end{equation}
where $n_{p\s} = d_{p\s}^{\dag} d_{p\s}^{\pdag}$ for $p = e$ or $o$. The
phonon-assisted tunneling component of $\tilde{H}_{\tun}$ (i.e., the original
$H_{\tun}$) can be eliminated by performing a second Lang-Firsov transformation
\begin{equation}
\hat{H}=e^{S_2}\tilde{H} e^{-S_2}
\end{equation}
with
\begin{equation}
\label{S_2}
S_2=\frac{\lambda'}{\hbar\omega_0}(n_e-n_o) \bigl(b^{\dag}-b\bigr),
\end{equation}
where $n_p=n_{p\up}+n_{p\dn}$.
Lengthy algebra reveals a transformed Hamiltonian
\begin{equation}
\label{hat H}
\hat{H} = \hat{H}_{\el} + H_{\ph} + H_{\leads} + \hat{H}_{\molleads},
\end{equation}
where
\begin{align}\label{hat H_el}
\hat{H}_{\el}
&=\sum_{p=e,o}\Bigl(\tilde{\e}_p \, n_p + \tilde{U}_p \, n_{p\up}n_{p\dn}\Bigr)
 +\sum_{\s} \Bigl(\tilde{U}_{\parallel} \, n_{e\s}n_{o\s}
 +\tilde{U}_{\!\perp} \, n_{e\s} n_{o\sbar} \Bigr) \notag \\
&+\sum_{\s} \bigl[t+W\,(n_{e\sbar}+n_{o\sbar})\bigr]
 \Bigl(B_{2\lambda'}^{\dag}d_{e\s}^{\dag} d_{o\s}^{\pdag}
 +B_{2\lambda'}^{\pdag}d_{o\s}^{\dag} d_{e\s}^{\pdag} \Bigr) \\
&+J\,\Bigl(S_e^+ S_o^- +S_o^+ S_e^-
 +B_{4\lambda'}^{\dag}I_e^+ I_o^- +B_{4\lambda'}^{\pdag}I_e^- I_o^+ \Bigr)
 \notag
\end{align}
with $S_p^+ \equiv \bigl(S_p^-\bigr)^{\dag} = c_{p\up}^{\dag} c_{p\dn}^{\pdag}$
and $I_p^+ \equiv \bigl(I_p^-\bigr)^{\dag} = c_{p\up}^{\dag} c_{p\dn}^{\dag}$
being spin- and charge-raising operators, respectively, and
\begin{equation}
\label{hat H_molleads}
\hat{H}_{\molleads}=\frac{2}{\sqrt{N_s}}\sum_{p=e,o}V_p\sum_{\bk,\s}
 \Bigl(B_{\lambda+\lambda'}^{\dag}d_{p\s}^{\dag}c_{e\bk\s}^{\pdag}
 +B_{\lambda+\lambda'}^{\pdag}c_{e\bk\s}^{\dag} d_{p\s}^{\pdag}\Bigr).
\end{equation}
The renormalized parameters entering Eqs.\ \eqref{hat H_el} and
\eqref{hat H_molleads} are
\begin{subequations}
\label{params}
\begin{align}
\tilde{\e}_p&=\frac{\e_\alpha+\e_\beta}{2}
 -\frac{\lambda_p^2}{\hbar\omega_0}, \\
\tilde{U}_p&=\frac{2U'+U_{\alpha}+U_{\beta}}{4}
 -\frac{2\lambda_p^2}{\hbar\omega_0}, \\
\tilde{U}_{\parallel}&=U'-\frac{2\lambda_e\lambda_o}{\hbar\omega_0}, \\
\tilde{U}_{\!\perp}&=\frac{2U'+U_{\alpha}+U_{\beta}}{4}
 -\frac{2\lambda_e\lambda_o}{\hbar\omega_0}, \\
t&=\frac{\e_{\alpha}-\e_{\beta}}{2}, \\
W&=\frac{U_{\alpha}-U_{\beta}}{2}, \\
J&=\frac{2U'-U_{\alpha}-U_{\beta}}{4}, \\
V_{e,o}&=\frac{V_{\alpha}\pm V_{\beta}}{2},
\end{align}
\end{subequations}
where
\begin{equation}
\label{lambda_p}
\lambda_{e,o} = \lambda \pm \lambda' .
\end{equation}
Since the e-e interactions are expressed much less compactly in the parity
basis than in the original basis of $\alpha$ and $\beta$ orbitals, the
elimination of the boson-assisted interorbital tunneling from the Hamiltonian
comes at the price of much greater complexity in $\hat{H}_{\el}$ compared to
$H_{\el}$ [Eq.\ \eqref{H_el}] and $\tilde{H}_{\el}$. It is notable, though,
that the e-e repulsion between two electrons within the even-parity
[odd-parity] molecular orbital undergoes a non-negative reduction proportional
to $\lambda_e^2=(\lambda+\lambda')^2$ [$\lambda_o^2=(\lambda-\lambda')^2$]. By
contrast, the Coulomb repulsion between electrons in orbitals of different
parity undergoes a shift proportional to
$-\lambda_e\lambda_o=\lambda'^2-\lambda^2$ that may be of either sign. Whereas
large values of $\lambda$ favor double occupancy of both the $\alpha$ and the
$\beta$ molecular orbital, large values of $|\lambda'|$ favor double occupancy
of either the $e$ or the $o$ linear combination [the degeneracy between these
alternatives being broken by an amount
$(2\tilde{\e}_e+\tilde{U}_e)-(2\tilde{\e}_o+\tilde{U}_o)
= -16\lambda\lambda'/\hbar\omega_0$].
Both limits yield a unique many-body ground state of a very different
character than the spin-singlet Kondo state.

Since $S_1$ defined in Eq.\ \eqref{S_1} can be rewritten
$S_1=(\lambda/\hbar\omega_0) (n_e+n_o) \bigl(b^{\dag}-b\bigr)$, it
commutes with $S_2$ given in Eq.\ \eqref{S_2}. As a result, the two
Lang-Firsov transformations can be combined into a single canonical
transformation
\begin{equation}
\hat{H}=e^{S}H e^{-S}
\end{equation}
with
\begin{equation}
\label{S}
S=S_1+S_2=\frac{\lambda_e n_e+\lambda_o n_o}{\hbar\omega_0}
\bigl(b^{\dag}-b\bigr) .
\end{equation}
This canonical transformation maps the original phonon annihilation operator
$b$ to
\begin{equation}
\label{hat b}
\hat{b}=e^S b e^{-S} = b
 - \frac{\lambda_e n_e+\lambda_o n_o}{\hbar\omega_0} .
\end{equation}
Since $\hat{b}^{\dag}-\hat{b}=b^{\dag}-b$, Eq.\ \eqref{B} can be rewritten
\begin{equation}
\label{hat B}
B_{\xi}=\exp\biggl[-\frac{\xi}{\hbar\omega_0}
 \bigl(\hat{b}^{\dag}-\hat{b}\bigr)\biggr] .
\end{equation}
Thus, the operators $B_{2\lambda'}$ and $B_{4\lambda'}$ entering Eq.\
\eqref{hat H_el}, as well as $B_{\lambda+\lambda'}$ in Eq.\
\eqref{hat H_molleads}, can be reinterpreted as leading to changes in the
occupation $\hat{n}_b\equiv\hat{b}^{\dag}\hat{b}$ of the transformed phonon
mode.

If the phonon energy $\hbar\omega_0$ were to greatly exceed the thermal
energy $k_B T$ and all other energy scales within the model, the system's
low-energy states would be characterized by $\langle\hat{n}_b\rangle\simeq 0$
or, equivalently,
\begin{equation}
\label{n_b:large-omega_0}
\langle n_b\rangle\equiv\langle b^{\dag}b\rangle \simeq\biggl\langle\biggl(
 \frac{\lambda_e n_e+\lambda_o n_o}{\hbar\omega_0}\biggr)^2\biggr\rangle.
\end{equation}
Moreover, one could approximate other physical quantities by taking
expectation values in the transformed phonon vacuum. This approach, which was
pioneered in the treatment of the small-polaron problem,\cite{Holstein:59a}
becomes exact in the anti-adiabatic limit $\omega_0\to\infty$. However, the
physical limit of greatest interest in the two-orbital molecule is one in which
the Coulomb interactions $U_{\alpha}$, $U_{\beta}$, and $U'$---and hence quite
possibly the couplings $|W|$ and $|J|$ associated with changes in
$\hat{n}_b$---are larger than $\hbar\omega_0$. The applicability to such
situations of the approximation $\hat{n}_b=0$, and of
Eq.\ \eqref{n_b:large-omega_0} in particular, is addressed in
Sec.\ \ref{sec:results}.

\section{Numerical renormalization-group approach}
\label{sec:method}

In order to obtain a robust description of the many-body physics of the model,
we treat the Hamiltonian \eqref{H} using Wilson's numerical
renormalization-group (NRG) method,\cite{RevModPhys.47.773,KWW:1980,
RevModPhys.80.395} as extended to incorporate local bosonic degrees of
freedom.\cite{Hewson:02}
The effective conduction band formed by the even-parity combination of left-
and right-lead electrons is divided into logarithmic bins spanning the energy
ranges $D\Lambda^{-(m+1)}<\pm\e<D\Lambda^{-m}$ for $m = 1$, $2$, $3$, $\ldots$,
for some discretization parameter $\Lambda>1$. After the continuum of band
states within each bin is approximated by a single representative state (the
linear combination of states within the bin that couples to the molecular
orbitals), Eq.\ \eqref{H_leads:mod} is mapped via a Lanczos transformation to
\begin{equation}
H_{\leads}\simeq\sum_{n=0}^{\infty} \sum_{\s} \tau_n \bigl( f^{\dag}_{n\s}
 f^{\pdag}_{n+1,\s} + f^{\dag}_{n+1,\s} f^{\pdag}_{n\s}\bigr) ,
\end{equation}
representing a semi-infinite, nearest-neighbor tight-binding chain to which
the impurity couples only at its end site $n=0$. Since the hopping decays
exponentially along the chain as $\tau_n\sim D\Lambda^{-n/2}$, the ground state
can be obtained via an iterative procedure in which iteration $N$ involves
diagonalization of a finite chain spanning sites $n\le N$. At the end of
iteration $N$, a pre-determined number of low-lying many-body eigenstates is
retained to form the basis for iteration $N+1$, thereby allowing reliable
access to the low-lying spectrum of chains containing tens or even hundreds of
sites. See Ref.\ \onlinecite{RevModPhys.80.395} for general details of the
NRG procedure.

For our problem, NRG iteration $N=0$ treats a Hamiltonian $H_0=H_{\mol}
+H_{\molleads}$, with $N_s^{-1/2}\sum_{\bk} c_{e\bk\s}$ in
Eq.\ \eqref{H_molleads:mod} replaced by $\sqrt{2} f_{0\s}$.
Since the phonon mode described by $H_{\ph}$ has an infinite-dimensional
Hilbert space, we must work in a truncated space in which the boson number is
restricted to $n_b \le N_b$.

\subsection{Thermodynamic quantities}
\label{subsec:thermodynamics}

The NRG method can be used to evaluate a thermodynamic property $X$ as
\begin{equation}
\label{X(T)}
X(T) = \frac{1}{Z(T)} \sum_m \langle \Psi_m|X|\Psi_m\rangle \:
 e^{-\beta E_m},
\end{equation}
where $|\Psi_m\rangle$ is a many-body eigenstate at iteration $N$ having
energy $E_m$, $\beta = 1/k_B T$, and
\begin{equation}
\label{Z(T)}
Z(T) = \sum_m e^{-\beta E_m}
\end{equation}
is the partition function evaluated at the same iteration. For a given value
of $N$, Eqs.\ \eqref{X(T)} and \eqref{Z(T)} provide a good
account\cite{RevModPhys.47.773,KWW:1980,RevModPhys.80.395} of $X(T)$ over a
range of temperatures around $T_N$ defined by $k_B T_N = D \Lambda^{-N/2}$.

For extensive properties $X$, it is useful to define the molecular contribution
to the property as
\begin{equation}
X_{\mol}=X_{\tot}-X_{\leads},
\end{equation}
where $X_{\tot}$ ($X_{\leads}$) is the total value of $X$ for a system
with (without) the molecule.
In our problem, the local phonon mode is treated as part of the host system.
Accordingly, we define the molecular entropy as
\begin{equation}
\label{without}
S_{\!\mol}(T)=S_{\!\tot}(T)-S_{\!\leads}(T)-S_{\!\ph}(T),
\end{equation}
where $S_{\!\tot}(T)$ is the total entropy of the system, $S_{\!\leads}(T)$ is
the contribution of the leads when isolated from the molecule, and
$S_{\!\ph}(T)$ is the entropy of the truncated local-phonon system, given by
\begin{equation}
S_{\!\ph}(T)=k_B\bigl[\ln Z_{\ph}(T) - \partial \ln Z_{\ph}/\partial\beta\bigr],
\end{equation}
with
\begin{equation}
Z_{\ph}(T)=\sum_{n_b=0}^{N_b} e^{-n_b\beta\hbar\omega_0}
=\frac{1-e^{-\beta \hbar \omega_0(N_b+1)}}{1-e^{-\beta \hbar \omega_0}}.
\end{equation}
Another property of interest is the molecular contribution to the static
magnetic susceptibility,
\begin{equation}
\chi(T)=\frac{\beta(g\mu_B)^2}{Z(T)}\sum_m
 \Bigl[\langle \Psi_m | S_z^2 |\Psi_m\rangle
 - |\langle \Psi_m | S_z |\Psi_m\rangle|^2\Bigr] e^{-\beta E_m},
\end{equation}
where $S_z$ is the total spin $z$ operator, $\mu_B$ is the Bohr magneton, and
$g$ is the Land\'e g factor (assumed to be the same for electrons in the leads
and in the molecular orbitals). One can interpret $|\bm{\mu}_{\mol}|^2 =
3 Tk_B\chi_{\mol}$ as the magnitude-squared of the molecule's effective
magnetic moment.

\subsection{Linear-response transport properties}
\label{subsec:linear response}

In this paper, we restrict our calculations to equilibrium situations in which
no external bias is applied. In such cases, inelastic transport produced by
the e-ph interaction can be neglected\cite{J.Phys.:Condens.Matter.4.5309} and
the linear conductance through the molecule can be obtained from a Landauer-type
formula
\begin{equation}
\label{G:general}
G(T)=G_0 \int_{-\infty}^{\infty}
 \biggl(-\frac{\partial f}{\partial\omega}\biggr)
 \: \bigl[ - \mathrm{Im} \, \mathcal{T}(\omega,T) \bigr] \: d\omega,
\end{equation}
where
\begin{equation}
\label{tau}
\mathcal{T}(\omega,T)=
\frac{\pi}{2D} \, \sum_{i=\alpha,\beta}\sum_{\s}V_{Li}\,
 G^{\,\s}_{ij}(\omega,T) \, V_{jR} .
\end{equation}
and $G_0=2e^2/h$ is the quantum of conductance. The fully dressed retarded
molecular Green's functions $G^{\,\s}_{ij}(\omega,T)$ are defined by
\begin{equation}
\label{G_ij}
G^{\,\s}_{ij}(\omega,T)=-i\int_0^\infty\Bigl\langle\Bigl[d_{i\s}^{\pdag}(t),
 d_{j\s}^{\dag}(0)\Bigr]_+\Bigr\rangle \, e^{i(\omega+i\eta)t} \, dt,
\end{equation}
where $\langle \cdots \rangle$ represents the equilibrium average in the
grand canonical ensemble and $\eta$ is a positive infinitesimal real number.

As shown for the related problem of two quantum dots connected in common to a
pair of metallic leads,\cite{Logan:09} in the case $V_{\ell i}=V_{i\ell}=V_i$
assumed in the present work, Eq.\ \eqref{G:general} can be recast in
the simpler form
\begin{equation}
\label{G}
G(T)/G_0 = \pi \, \Gamma_c \sum_{\s} \int_{-\infty}^{\infty}
 \biggl(-\frac{\partial f}{\partial\omega}\biggr)
 \: A_{cc}^{\,\s}(\omega,T)
 \: d\omega,
\end{equation}
where $\Gamma_c=\Gamma_{\alpha}+\Gamma_{\beta}$ and
$A_{cc}^{\,\s}(\omega,T)=\pi^{-1}\mathrm{Im}\,G^{\,\s}_{cc}(\omega,T)$ [defined
via Eq.\ \eqref{G_ij}] is the spectral function for the current-carrying linear
combination of the $\alpha$ and $\beta$ orbitals:
\begin{equation}
d_{c\s} = \sum_{i=\alpha,\beta} \sqrt{\Gamma_i/\Gamma_c} \, d_{i\s} .
\end{equation}
Within the NRG approach, one can calculate
\begin{align}\label{spectral}
A_{cc}^{\s}(\omega,T)
& =\frac{1}{Z}\sum_{m,m'} \bigl|\langle\Psi_{m'}|d_{c\s}^{\dag}
  |\Psi_{m}\rangle\bigr|^2 \Bigl(e^{-\beta E_m}+e^{-\beta E_{n'}}\Bigr)
  \notag \\
& \quad \times \,
  \delta_T\bigl(\omega-(E_{m'}-E_m)/\hbar\bigr) ,
\end{align}
where $\delta_T(\omega)$ is a thermally broadened Dirac delta
function.\cite{RevModPhys.80.395} We consider only situations where there is no
magnetic field, and hence $A_{cc}^{\s}(\omega, T)=A_{cc}(\omega, T)$
independent of $\s$.

\section{Results}
\label{sec:results}

This section presents and interprets essentially exact NRG results for the
Hamiltonian defined by Eqs.\ \eqref{H}--\eqref{H_tun}, \eqref{H_molleads:mod},
and \eqref{H_leads:mod}. We have been guided in our choice of model parameters
by the physical considerations laid out at the end of Sec.\ \ref{subsec:model}.
We take the half bandwidth $D=1$ as our primary energy scale and adopt units
in which $\hbar=k_B=g\mu_B=1$.

The results shown below were all obtained for the special case of equal orbital
hybridizations $V_{\alpha}=V_{\beta}=V$ and equal intraorbital Coulomb
repulsions $U_{\alpha}=U_{\beta}=U$. These choices, which simplify algebraic
analysis because they lead to $W=0$ in Eq.\ \eqref{hat H_el} and $V_o=0$ in
Eq.\ \eqref{hat H_molleads}, are not crucial; qualitatively very similar
results are obtained in more general cases. Most of the numerical data were
computed for equal intraorbital and interorbital interactions $U=U'=0.5$.
However, we also include results for other values of $U'/U$ and for the
limiting cases $U=U'=0$ and $U=U'=5$.

Our calculations were performed for phonon energy $\omega_0=0.1$ and
hybridization $V=0.075$, resulting in an orbital width
$\Gamma=\pi V^2/D \simeq 0.0177$. As discussed in Sec.\ \ref{subsec:model},
the resulting ratio $\omega_0/\Gamma\simeq 6$ places the system in the
anti-adiabatic regime of greatest interest from the perspective of
competition between e-e and e-ph effects. For this fixed value of
$\omega_0/\Gamma$, we show the consequences of changing the e-ph
couplings (a variation of theoretical interest that may be impractical in
experiments) and the orbital energies (which can likely be achieved by tuning
gate voltages).

Finally, all calculations were performed using an NRG discretization parameter
$\Lambda=2.5$, allowing up to $N_b=60$ phonons in the local mode, and retaining
2\,000--4\,000 many-body states after each iteration. These choices are
sufficient to reduce NRG discretization and truncation errors to minimal levels.

\subsection{Large orbital energy separation $\e_{\beta}-\e_{\alpha}$}
\label{subsec:large-e_b}

We first consider the case of fixed $\e_{\beta}=4$ where the upper molecular
orbital lies far above the chemical potential of the leads and therefore
contributes little to the low-energy physics. This situation, in which the
two-orbital model largely reduces to the Anderson-Holstein
model,\cite{Simanek:79,Ting:80,Kaga:80,Schonhammer:84,Alascio:88,Schuttler:88,
Ostreich:91,Hewson:02,Jeon:03,Zhu:03,Lee:04,Cornaglia:04,Cornaglia:05} serves
as a benchmark against which to compare cases in which both molecular orbitals
are active.

Given that the $\beta$ orbital will have negligible occupation, the
interorbital Coulomb repulsion $U'$ entering $H_{\el}$ [Eq.\ \eqref{H_el}]
and the interorbital e-ph coupling $\lambda'$ entering $H_{\tun}$ [Eq.\
\eqref{H_tun}] are not expected to greatly affect the low-energy properties.
Throughout this subsection we assume $U'=U$ to reduce the number of different
parameters that must be specified. Figures
\ref{large-e_b occs}--\ref{large-e_b props} present results obtained for
$\lambda'=\lambda$; switching to $\lambda'=-\lambda$ would interchange the
roles of the even and odd linear combinations of molecular orbitals, but would
not change any of the physical quantities shown. Figures
\ref{large-e_b compare Tstar} and \ref{large-e_b compare G} demonstrate that
very similar properties arise for $\lambda'=0$.

\subsubsection{Isolated molecule}
\label{subsubsec:large-e_b isolated}

We begin by using the transformed Hamiltonian $\tilde{H}$ defined in
Eq.\ \eqref{tilde H} to find analytical expressions for the energies of the
low-lying states of the isolated molecule in the absence of any electron
tunneling to/from the leads (i.e., for $V=0$). In the regime where
$\tilde{\e}_{\beta}$ is the largest energy scale of the molecule, $\lambda'$
manifests itself primarily through perturbative corrections to the energies of
the molecule when the $\alpha$ orbital is occupied by $n_{\mol}=0$, $1$, or $2$
electrons.

Let us focus on the state of lowest energy in each occupancy sector. This
is the state having zero occupancy of the transformed boson mode entering the
Hamiltonian $\tilde{H}$, whose energy we will denote $E^{(n_{\mol})}_{\min}$.
The empty molecule is unaffected by the interorbital e-ph coupling, so
$E^{(0)}_{\min}=0$. To second order in $\tilde{H}_{\tun}$ defined in
Eq.\ \eqref{tilde H_tun},
\begin{align}
\label{tilde tilde e_a}
E^{(1)}_{\min}
\equiv \tilde{\tilde{\e}}_{\alpha}
&= \tilde{\e}_{\alpha}
   - \frac{\lambda'^2}{\tilde{\e}_{\beta}-\tilde{\e}_{\alpha}+\omega_0}
   - \biggl(\frac{2\lambda}{\omega_0}\biggr)^2
     \frac{\lambda'^2}{\tilde{\e}_{\beta}-\tilde{\e}_{\alpha}} \notag \\
&\simeq \e_{\alpha} - \lambda^2/\omega_0
 - ( 1 + 4\lambda^2/\omega_0^2) \: \lambda'^2/\e_{\beta} ,
\end{align}
where in the second expression we have used
$\tilde{\e}_{\beta}-\tilde{\e}_{\alpha} = \e_{\beta}-\e_{\alpha}$.
In the same approximation, the energy of the doubly occupied molecule becomes
\begin{align}
\label{E^2_min}
E^{(2)}_{\min}
&= 2\tilde{\e}_{\alpha} + \tilde{U}_{\alpha}
  - \frac{2\lambda'^2}{\tilde{\e}_{\beta}-\tilde{\e}_{\alpha}+\tilde{U}'
  -\tilde{U}_{\alpha}+\omega_0} \notag \\
&\qquad - \biggl(\frac{4\lambda}{\omega_0}\biggr)^2
    \frac{2\lambda'^2}{\tilde{\e}_{\beta}-\tilde{\e}_{\alpha}+\tilde{U}'
  -\tilde{U}_{\alpha}} \notag \\
&\simeq 2\tilde{\e}_{\alpha} + \tilde{U}
  - 2 ( 1 + 16\lambda^2/\omega_0^2 ) \: \lambda'^2/\e_{\beta}
\end{align}
in the case $U_{\alpha}=U'=U$ considered throughout this discussion of large
orbital separation. Equations \eqref{tilde tilde e_a} and \eqref{E^2_min}
allow us to define an effective interaction within the $\alpha$ orbital
\begin{equation}
\label{tilde tilde U_a}
\tilde{\tilde{U}}_{\alpha} = E^{(2)}_{\min} - 2 E^{(1)}_{\min}
= U - 2\lambda^2/\omega_0
  - 24 (\lambda^2/\omega_0^2) \: \lambda'^2 / \e_{\beta} .
\end{equation}
For future reference, we also define
\begin{equation}
\label{Delta E_12}
\Delta E_{12} = E^{(1)}_{\min}-E^{(2)}_{\min}
= 3\lambda^2/\omega_0 + (1+28\lambda^2/\omega_0^2)
\, \lambda'^2/\e_{\beta}-\e_{\alpha}-U .
\end{equation}

The ground state of the isolated molecule lies in the sector of
occupancy $n_{\mol}$ having the smallest value of $E^{(n_{\mol})}_{\min}$.
Under variation of a molecular parameter such as $\e_{\alpha}$ or $\lambda$,
a jump will occur between $n_{\mol}=0$ and 1 at any point where
$E^{(1)}_{\min}=0<E^{(2)}_{\min}$, between $n_{\mol}=1$ and 2 where
$E^{(2)}_{\min}=E^{(1)}_{\min}<0$, and directly between $n_{\mol}=0$ and 2
where $E^{(2)}_{\min}=0<E^{(1)}_{\min}$. In the presence of a small level width
$\Gamma>0$, one expects these jumps to be broadened into smooth crossovers
centered at points in parameter space close to their locations for the isolated
molecule.

\subsubsection{Effect of varying the lower orbital energy}
\label{subsubsec:large-e_b sweeps}

Now we turn to numerical solutions of the full problem with
$\e_{\beta}=4$, a dot-lead hybridization $V=0.075$, and a phonon energy
$\omega_0=0.1$. In this subsection we examine the effect of varying the energy
$\e_{\alpha}$ of the lower molecular orbital at $T=0$.

\begin{figure}[tb]
\centering\includegraphics[width=3.25in]{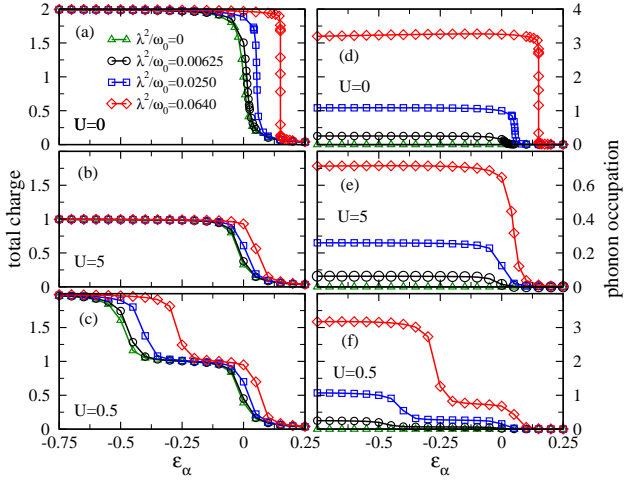}
\caption{\label{large-e_b occs} (Color online)
Variation with orbital energy $\e_\alpha$ at zero temperature of (left) the
charge $\langle n_{\mol}\rangle$, and (right) the phonon occupation
$\langle n_b\rangle$, for $U=0$ (top), $U=5$ (middle) and $U=0.5$ (bottom).
Data are for $\e_{\beta}=4$, $U'=U$, $\lambda'=\lambda$, and the four values
of $\lambda^2/\omega_0$ listed in the legend.}
\end{figure}

Figure \ref{large-e_b occs} shows the total molecular charge
$\langle n_{\mol}\rangle$ and the occupation $\langle n_b\rangle$ of the
original phonon mode [as opposed to the occupation $\langle \hat{n}_b\rangle$
of the transformed mode defined in Eq.\ \eqref{hat b}] as functions of
$\e_\alpha$ for four values of $\lambda$ and three values of $U$. First consider
the case $U=0$ of vanishing e-e interactions shown in panels (a) and (d). For
$\lambda=\lambda'=0$, $\e_{\alpha}=0$ is a point of degeneracy between
configurations having molecular charges 0, 1, and 2; $\langle n_{\mol}\rangle$
increases from 0 to 2 over a narrow range $\Delta\e_{\alpha}\simeq 4\Gamma$ as
the $\alpha$ orbital drops below the chemical potential of the leads.
For $\lambda>0$, $\tilde{U}=-2\lambda^2/\omega_0$ is negative, and the
ground state switches from charge 0 to charge 2 around the point where
$E^{(2)}_{\min}=E^{(0)}_{\min}$ or $\e_{\alpha} = 2\lambda^2 / \omega_0
+ (1+16\lambda^2/\omega_0^2)\,\lambda'^2/\e_{\beta}$.
There is a marked decrease with increasing $\lambda$ in the width
$\Delta\e_{\alpha}$ of the region of rapid change in the charge. (We will
henceforth refer to such a measure as the ``rise width'' to avoid possible
confusion with the width of the plateau between two successive rises.)

It is evident from Figs.\ \ref{large-e_b occs}(a) and \ref{large-e_b occs}(d)
that changes in the ground-state phonon occupation are closely correlated
with those in the total molecular charge. The prediction of Eq.\
\eqref{n_b:large-omega_0} for the case $\lambda'=\lambda$ (hence,
$\lambda_e=2\lambda$ and $\lambda_o=0$) is $\langle n_b\rangle =
(2\lambda/\omega_0)^2 \, \langle n_e^2\rangle$.
Although this relation captures the correct trends in the variation of
$\langle n_b\rangle$ with $\e_{\alpha}$ in Fig.\ \ref{large-e_b occs}(d), it
overestimates the phonon occupation by a significant margin. Such deviations
are not unexpected, given that Eq.\ \eqref{n_b:large-omega_0} was derived under
the assumption that $\hbar\omega_0$ is the largest energy scale in the problem,
whereas here $\e_{\beta}$ is the dominant energy scale, followed by $U=U'$.
Empirically, we find that $\langle n_b\rangle$ lies closer to
\begin{equation}
\label{nbar_b}
  \bar{n}_b = (2\lambda/\omega_0)^2 \, \langle n_e\rangle^2,
\end{equation}
which also serves as an empirical lower bound on the phonon occupation. The
error $\langle n_b\rangle-\bar{n}_b$ is largest in the vicinity of the sharpest
rise in $\langle n_{\mol}\rangle$ and vanishes as $\langle n_{\mol}\rangle$
approaches 0 or 2.
For $\lambda^2/\omega_0=0.064$, $\langle n_b\rangle-\bar{n}_b < 0.06$
both for $\e_{\alpha}\le 0.149$ and for $\e_{\alpha}\ge 0.151$, whereas for
$\e_{\alpha}=0.149945$, $\bar{n}_b\simeq 0.88$ underestimates
$\langle n_b\rangle$ by approximately $0.68$. The peak error is smaller for
the other e-ph couplings shown in Fig.\ \ref{large-e_b occs}(d).

For the case $U=5$ of very strong e-e interactions [Fig.~\ref{large-e_b occs}(b)
and \ref{large-e_b occs}(e)], the molecular charge rises from 0 to 1 around the
point where $E^{(1)}_{\min}=E^{(0)}_{\min}$ or $\e_\alpha \simeq
2\lambda^2/\omega_0 + (1+4\lambda^2/\omega_0^2) \, \lambda'^2/\e_{\beta}$.
In contrast with the situation for $U=0$, the rise width $\Delta\e_{\alpha}$
shows no appreciable change with $\lambda$. The phonon occupation is described
by Eq.\ \eqref{nbar_b} even better than for $U=0$, with the greatest error
($\langle n_b\rangle - \bar{n}_b\simeq 0.08$ for $\lambda^2/\omega_0=0.064$)
occurring around the point where $\langle n_{\mol}\rangle=0.5$.

Lastly, Figs.~\ref{large-e_b occs}(c) and \ref{large-e_b occs}(f) show data for
$U=0.5$, exemplifying moderately strong e-e interactions. With decreasing
$\e_{\alpha}$ (at fixed $\lambda$), the molecular charge rises in two steps,
first rising from 0 to 1 as $E^{(1)}_{\min}$ falls below $E^{(0)}_{\min}$,
and then rising from 1 to 2 as $E^{(1)}_{\min}$ in turn falls below
$E^{(2)}_{\min}$ at [see Eq.\ \eqref{Delta E_12}]
\begin{equation}
\e_\alpha \simeq -U + 3\lambda^2/\omega_0 + (1+28\lambda^2/\omega_0^2) \,
\lambda'^2/\e_{\beta}.
\end{equation}
Just as for $U=5$, each rise has a width $\Delta\e_{\alpha}=O(\Gamma)$ that is
independent of $\lambda$ over the range of e-ph couplings shown. The distance
along the $\e_{\alpha}$ axis between the two rises (i.e., the width of the
charge-1 plateau) is roughly $\tilde{\tilde{U}}$ defined in Eq.\
\eqref{tilde tilde U_a}, which decreases as the e-ph coupling increases in
magnitude. The phonon occupation is again well-approximated by $\bar{n}_b$
given in Eq.\ \eqref{nbar_b}.

\begin{figure}[tb]
\centering\includegraphics[width=3in]{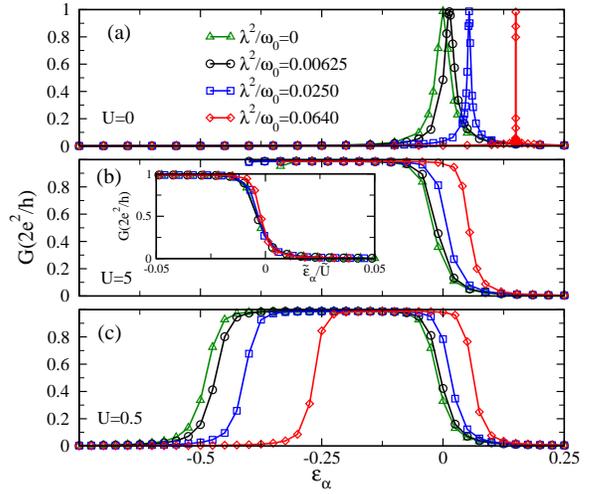}
\caption{\label{large-e_b cond} (Color online)
Zero-temperature conductance $G$ vs orbital energy $\e_\alpha$ for (a) $U=0$,
(b) $U=5$, and (c) $U=0.5$. Data are for $\e_{\beta}=4$, $U'=U$,
$\lambda'=\lambda$, and the four values of $\lambda^2/\omega_0$ listed in the
legend. The inset in (b) shows the data from the main panel replotted as $G$
vs $\tilde{\e}_\alpha/\tilde U$.}
\end{figure}

Figure \ref{large-e_b cond} plots the zero-temperature linear conductance as a
function of $\e_\alpha$ for the same set of parameters as was used in Fig.\
\ref{large-e_b occs}. At $T=0$ in zero magnetic field, Eq.\ \eqref{G} reduces
to $G(T=0)/G_0=\pi\,\Gamma_c A_{cc}(0,0)$. In any regime of Fermi-liquid
behavior, $A_{cc}(0,0)$ is expected to obey the Friedel sum-rule, implying
that $\pi\,\Gamma_c A_{cc}(0,0)=\sin^2(\pi\langle n_{\mol}\rangle/2)$ in the
wide-band limit where all other energy scales in the model are small compared
with $D$. This property, which should hold even in the presence of e-ph
interactions within the molecule, leads to
\begin{equation}
\label{FL cond}
G(T=0)=G_0\,\sin^2\biggl(\frac{\pi}{2}\langle n_{\mol}\rangle\biggr)\,.
\end{equation}

For $U=0$ [Fig.\ \ref{large-e_b cond}(a)], we observe a conductance peak
at the point of degeneracy between molecular charges 0 and 2. This is the
noninteracting analog of the Coulomb blockade peak seen in strongly interacting
quantum dots and single-molecule junctions above their Kondo temperatures. For
$\lambda=0$, the peak is located at $\e_{\alpha}=0$ and has a full width
$\Delta\e_{\alpha}\simeq 2\Gamma$, as expected for this exactly solvable
single-particle case. With increasing $\lambda$, the conductance peak
shifts to higher $\e_\alpha$ while its width narrows, trends that both follow
via Eq.\ \eqref{FL cond} from the behavior of $\langle n_{\mol}\rangle$ in
Fig.\ \ref{large-e_b occs}(a). For all values of $\lambda$, the maximum
conductance is $G=G_0$, as predicted by Eq.\ \eqref{FL cond} for the point
where $\langle n_{\mol}\rangle$ passes through $1$.

The sharp features shown in Fig.\ \ref{large-e_b cond}(a) allow one to
quantify the accuracy of the approximation of using energies of the isolated
molecule in the no-boson state of the transformed phonon mode to locate
features in the full system. In the case $\lambda^2/\omega_0=0.064$, for
example, the NRG calculations place the peak in $G$ at $\e_{\alpha}=0.152$,
whereas the criterion $E^{(2)}_{\min}=E^{(0)}_{\min}$ gives $\e_\alpha =
2\lambda^2/\omega_0+(1+16\lambda^2/\omega_0^2)\,\lambda'^2/\e_{\beta} = 0.162$.
Thus, the coupling of the molecule to external leads and the admixture of states
with nonzero phonon number produces a downward shift in the peak position of
roughly $0.1$, considerably larger than the upward shift $0.018$ predicted to
arise from the $\lambda'$ interorbital e-ph coupling $\lambda'$.

For the interacting cases shown in Figs.\ \ref{large-e_b cond}(b) and
\ref{large-e_b cond}(c), the formation of a many-body Kondo resonance at the
chemical potential leads to a near-unitary conductance at low-temperatures
$T\ll T_K$ over the entire range of $\e_{\alpha}$ for which
$\langle n_{\mol}\rangle\simeq 1$. In the case $U=5$, no data are shown for
$\e_{\alpha} \lesssim -0.4$, a range in which the Kondo temperature $T_K$ is so
low that the ground-state properties are experimentally inaccessible.
For both nonzero values of $U$, the width of each conductance rise is
independent of $\lambda$ over the range of e-ph couplings shown.

The narrowing with increasing $\lambda$ of the rises in the molecular
charge and the phonon occupancy, and of the peaks in the linear conductance,
seen for $U=0$ but not in the data presented for $U=0.5$ or $U=5$, is associated
with the presence of a crossover of $\langle n_{\mol}\rangle$ directly from 0
to 2. Similar narrowing is, in fact, seen for $U>0$ when $\lambda$ become
sufficiently large to suppress the $\langle n_{\mol}\rangle=1$ plateau. (In
the case $U=0.5$ and $\lambda'=\lambda$, this takes place around
$\lambda^2/\omega_0=0.15$, considerably larger than any of the values shown in
Figs.\ \ref{large-e_b occs} and \ref{large-e_b cond}.) This phenomenon is
known from the Anderson-Holstein model (e.g., see Ref.\
\onlinecite{Cornaglia:04}) to arise from the small overlap between the bosonic
ground state of the displaced oscillator that minimizes the energy in the
sectors $n_{\mol}=0$ and the corresponding ground state for $n_{\mol}=2$.
This small overlap leads to an exponential reduction in the effective value of
the level width $\Gamma$ in the regime of negative effective $U$.

\begin{figure}[tb]
\centering\includegraphics[width=3in]{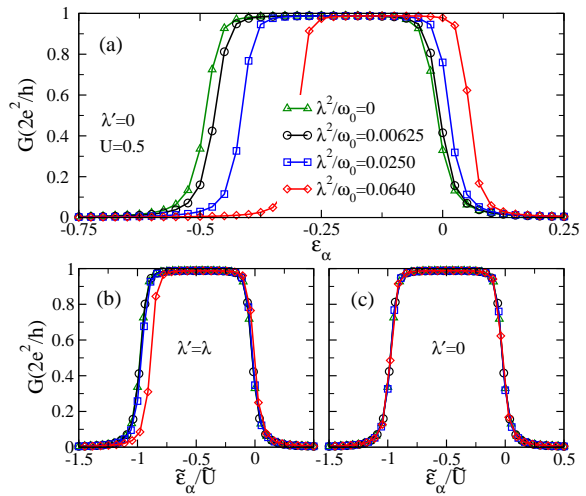}
\caption{\label{large-e_b compare G} (Color online)
(a) Zero-temperature conductance $G$ vs orbital energy $\e_\alpha$ for the
same parameters as used in Fig.\ \ref{large-e_b cond}(c), except
here $\lambda'=0$.
(b) Data for $\lambda'=\lambda$ from Fig.\ \ref{large-e_b cond}(c)
replotted as $G$ vs $\tilde{\e}_\alpha/\tilde U$.
(c) Data for $\lambda'=0$ from (a) replotted as $G$ vs
$\tilde{\e}_\alpha/\tilde U$.}
\end{figure}

It has already been remarked that the phonon-assisted interorbital
tunneling is expected to play only a minor role in cases where the $\beta$
orbital is far above the Fermi energy. To test this expectation, we have
compared data for $\lambda'=\lambda$ and $\lambda'=0$ with all other parameters
the same. The conductance curves in the two cases are also similar, as
exemplified for $U=0.5$ by Figs.\ \ref{large-e_b cond}(c) and
\ref{large-e_b compare G}(a). The same conclusion holds for the molecular
charge and phonon occupation (data for $\lambda'=0$ not shown). However, there
are subtle differences that can be highlighted by replotting properties as
functions of the scaling variable $\tilde{\e}_{\alpha}/\tilde U$. For example,
the conductance data for $\lambda'=0$ and $U=0.5$ show almost perfect collapse
[Fig.\ \ref{large-e_b compare G}(c)], confirming that in this case the
conductance rises are centered close to $\tilde{\e}_{\alpha}=0$ and
$\tilde{\e}_{\alpha}= -\tilde{U}$, the values predicted based on the low-lying
levels of the isolated molecule. For $\lambda'=\lambda$, the data collapse
[shown in Fig.\ \ref{large-e_b compare G}(b) for $U=0.5$, and in the inset to
Fig. \ref{large-e_b cond}(b) for $U=5$] is good for small values of $\lambda$
but less so for $\lambda^2/\omega_0=0.064$, a case where
$\tilde{\tilde{\e}}_{\alpha}$ and $\tilde{\tilde{U}}_{\alpha}$ defined in Eqs.\
\eqref{tilde tilde e_a} and \eqref{tilde tilde U_a} differ
appreciably from $\tilde{\e}_{\alpha}$ and $\tilde{U}$.

\subsubsection{Lower orbital close to chemical potential}
\label{subsubsec:large-e_b small-e_a}

\begin{figure}[tb]
\centering\includegraphics[width=3.25in]{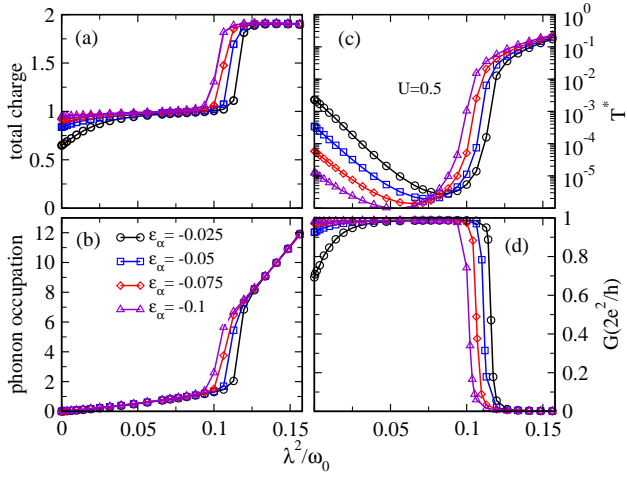}
\caption{\label{large-e_b props} (Color online)
Variation with $\lambda^2/\omega_0$ of (a) the ground-state molecular charge
$\langle n_{\mol}\rangle$, (b) the ground-state phonon occupation
$\langle n_b\rangle$, (c) the crossover temperature $T^*$, and (d) the
zero-temperature linear conductance $G$, all calculated for $U'=U=0.5$,
$\lambda'=\lambda$, $\e_{\beta}=4$, and the four values of $\e_{\alpha}$
listed in the legend.}
\end{figure}

We now switch focus from the variation of properties with $\e_{\alpha}$
to trends with increasing e-ph coupling. Figure \ref{large-e_b props}(a) shows
the evolution of the zero-temperature molecular charge $\langle n_{\mol}\rangle$
with $\lambda^2/\omega_0$ for $U=U'=0.5$, $\lambda'=\lambda$, and four different
values of $\e_{\alpha}$. We begin by considering the special case $\lambda=0$
in which the electron and phonon subsystems are entirely decoupled. Here,
$\langle n_{\mol}\rangle$ ranges from roughly two-thirds for
$\e_{\alpha}=-0.025$ (an example of mixed valence where the lower molecular
orbital lies below the Fermi energy by an amount that barely exceeds
$\Gamma=\pi V^2/D\simeq 0.0177$) to nearly one for $\e_{\alpha}=-0.075$
and $-0.1$. In the latter limit, the large Coulomb repulsion $U$ leads to
local-moment formation in the $\alpha$ orbital. The local moment is
collectively quenched by lead electrons, leading to a Kondo singlet ground
state. Figure \ref{large-e_b props}(c) shows the characteristic temperature
$T^*$ of the quenching of the molecular spin degree of freedom, determined via
the standard criterion\cite{RevModPhys.47.773} $T^*\chi_{\mol}(T^*)=0.0701$.
This scale is of order $\Gamma$ deep in the mixed-valence limit (i.e., for
$|\e_{\alpha}|\ll\Gamma$), but is exponentially reduced in the local-moment
regime $-\e_{\alpha}\gg\Gamma$ where it represents the system's Kondo
temperature, given for $U\gg-\e_{\alpha}\gg\Gamma$ by\cite{Haldane:78}
\begin{equation}
\label{T_K}
T_K\simeq\sqrt{\Gamma U} \, \exp(\pi\e_{\alpha}/2\Gamma).
\end{equation}

Upon initial increase of $\lambda$, the effective level position
$\tilde{\tilde{\e}}_{\alpha}$ decreases according to Eq.\
\eqref{tilde tilde e_a}, the occupancy of the lower molecular orbital (and
hence the total occupancy $\langle n_{\mol}\rangle$) rises ever closer to one,
and the temperature $T^*$ decreases as expected from the replacement of
$\e_{\alpha}$ and $U$ in Eq.\ \eqref{T_K} by $\tilde{\tilde{\e}}_{\alpha}$ and
$\tilde{\tilde{U}}_{\alpha}$. Neglecting both the subleading $\lambda$
dependence coming from $\tilde{U}$ and the $\lambda'$ contributions to
$\tilde{\tilde{\e}}_{\alpha}$, one arrives at the relation
\begin{equation}
T^*(\lambda)\simeq T^*(0)\:\exp\left[-\pi\lambda^2/(2\Gamma\omega_0)\right],
\end{equation}
which accounts quite well for the initial variation of $T^*$ in
Fig.\ \ref{large-e_b props}(c).

Upon further increase in the e-ph coupling, $\langle n_{\mol}\rangle$
and $T^*$ both show rapid but continuous rises around some value
$\lambda=\lambda_x$ that is close to the one predicted by the vanishing of
$E^{(2)}_{\min}=E^{(1)}_{\min}$ for the increase from 1 to 2 in the charge of
the isolated molecule: solving Eq.\ \eqref{Delta E_12} with
$\Delta E_{12}=0$ to find $\lambda=\lambda'$ yields
$\lambda_x^2/\omega_0=0.122$, $0.117$, $0.112$, and $0.106$ for
$\e_{\alpha} = -0.025$, $-0.05$, $-0.075$, and $-0.1$, respectively---values
close to but slightly above those observed in the full numerical solutions [the
magnitude and sign of the small discrepancies being consistent with those noted
previously in connection with the $U=0$ data in Fig.\ \ref{large-e_b cond}(a)].
For $\Gamma>0$, the energies corresponding to $E_1^{(1)}$ and $E_1^{(2)}$ each
acquire a half width $\Gamma$, so the crossover of the ground-state molecular
charge from 1 to 2 is smeared over the range
$|E^{(2)}_{\min}-E^{(1)}_{\min}| \lesssim 2\Gamma$. Solving Eq.\
\eqref{Delta E_12} again with $\Delta E_{12}=\pm 2\Gamma$ gives the full width
for the crossover as $\Delta(\lambda^2/\omega_0)\simeq 0.016$, an estimate in
good agreement with the data in Fig.\ \ref{large-e_b props}(a).

In the regime $\lambda\gtrsim\lambda_x$, minimization of the e-ph energy
through $\langle n_{\alpha}\rangle\simeq 2$, $\langle n_{\beta}\rangle\simeq 0$
outweighs the benefits of forming a many-body Kondo singlet. Therefore, $T^*$
characterizing the vanishing of $T\chi_{\mol}$ ceases to represent the Kondo
temperature and instead characterizes the scale, of order $\Delta E_{12}$
defined in Eq.\ \eqref{Delta E_12}, at which $n_{\mol}=1$ spin-doublet
molecular states become thermally inaccessible.

Over the entire range of $\delta$ and $\lambda^2/\omega_0$ illustrated in
Fig.\ \ref{large-e_b props}, the ground-state phonon occupation
$\langle n_b\rangle$ [Fig.\ \ref{large-e_b props}(b)] closely tracks $\bar{n}_b$
defined in Eq.\ \eqref{nbar_b} to within an absolute error
$0\le \langle n_b\rangle-\bar{n}_b\le 0.2$, an error that peaks around
$\lambda=\lambda_x$. Similarly, the $T=0$ conductance [Fig.\
\ref{large-e_b props}(d)] is everywhere well-described by Eq.\ \eqref{FL cond},
reaching the unitary limit $G_0$ over a window of Kondo behavior for
$\lambda\lesssim\lambda_x$ in which the molecular charge is 1, then plunging to
zero as the Kondo effect is destroyed and the occupancy rises to 2.

\begin{figure}[tb]
\centering\includegraphics[width=3.25in]{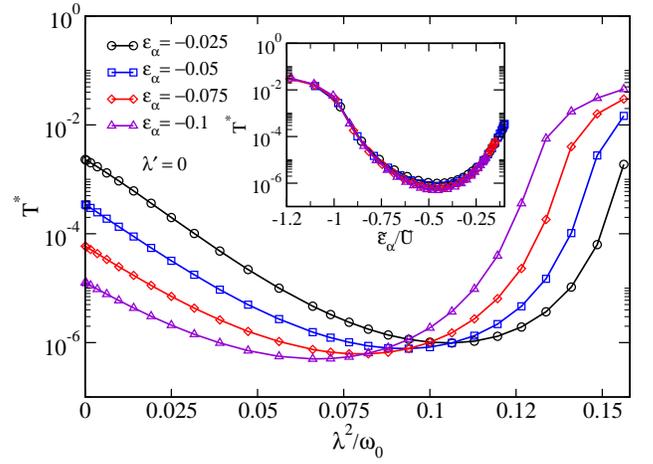}
\caption{\label{large-e_b compare Tstar} (Color online)
Variation with $\lambda^2/\omega_0$ of the crossover temperature $T^*$ calculated
for $U'=U=0.5$, $\lambda'=0$, $\e_{\beta}=4$, and the four values of $\e_{\alpha}$
listed in the legend. Inset: The same $T^*$ data plotted vs the ratio
$\tilde{\e}_{\alpha}/\tilde{U}_{\alpha}$ of phonon-renormalized molecular
parameters.}
\end{figure}

As another illustration of the effect of relaxing the assumption
$\lambda'=\lambda$, Fig.\ \ref{large-e_b compare Tstar} shows the variation with
$\lambda^2/\omega_0$ of $T^*$, calculated for the same parameters as in Fig.\
\ref{large-e_b props}(c), except that here $\lambda'=0$. For each value of
$\e_{\alpha}$, the variation of $T^*$ is very similar in the two cases apart
from a considerably larger value of $\lambda_x$ for $\lambda'=0$, a change that
is predicted at the level of the isolated molecule where Eq.\ \eqref{Delta E_12}
with $\lambda'=\Delta E_{12}=0$ gives $\lambda_x^2/\omega_0 = (U+\e_{\alpha})/3$,
which ranges from $0.158$ for $\e_{\alpha}=-0.025$ to $0.133$ for
$\e_{\alpha}=-0.1$. Just as seen in Fig.\ \ref{large-e_b compare G}(c), the
$\lambda'=0$ data exhibit excellent collapse when plotted against the ratio
$\tilde{\e}_{\alpha}/\tilde{U}$ of effective molecular parameters defined in
Eqs.\ \eqref{tilde quantities}.

\subsection{Small orbital energy separation $\e_{\beta}-\e_{\alpha}$}
\label{subsec:small-delta}

The rich behavior of the model described by Eqs.\ \eqref{H}--\eqref{H_molleads}
becomes apparent only in the regime where the two molecular orbitals lie close
in energy so that they can both contribute strongly to the low-energy physics.
For simplicity, we focus primarily on situations with equal e-ph couplings
$\lambda'=\lambda$, equal Coulomb interactions $U'=U$, and symmetrical
placement of the orbitals with respect to the chemical potential of the leads,
i.e., $\e_{\beta} = -\e_{\alpha} = \delta$, a small positive energy scale.
However, we present results for more general parameter choices at several points
throughout the subsection.

\subsubsection{Isolated molecule}
\label{subsubsec:small-delta isolated}

\begin{table}[tb]
\begin{center}
\setlength{\extrarowheight}{1ex}
\begin{tabular}{cc@{\hspace{2ex}}l@{\hspace{2ex}}l@{\hspace{2ex}}l}
$n_{\mol}$ & $i$
 &\multicolumn{1}{c}{$|\phi^{(n_{\mol})}_i(\delta\!=\!0)\rangle$
 \hspace{2em}\mbox{}}
 &$E^{(n_{\mol})}_i(\delta\!=\!0)$
 &$E^{(n_{\mol})}_i(\delta\!>\!0)$ \\[.5ex]
\hline
$0$ &$1$ &$|\hat{0}\rangle$ &$0$ &$0$ \\[.5ex]
\hline
$1$
&$1$ &$d_{e\up}^{\dag}|\hat{0}\rangle$
 &$-(x\!+\!x')^2$ &$-4x^2\!-\!\frac{1}{4}y$ \\
&$2$ &$d_{e\dn}^{\dag}|\hat{0}\rangle$
 &$-(x\!+\!x')^2$ &$-4x^2\!-\!\frac{1}{4}y$ \\
&$3$ &$d_{o\up}^{\dag}|\hat{0}\rangle$
 &$-(x\!-\!x')^2$ &$\frac{1}{4}y$ \\
&$4$ &$d_{o\dn}^{\dag}|\hat{0}\rangle$
 &$-(x\!-\!x')^2$ &$\frac{1}{4}y$ \\[.5ex]
\hline
$2$
&$1$ &$\bigl(c_1 d_{e\up}^{\dag}d_{e\dn}^{\dag}
   \!\!+\!c_2 d_{o\up}^{\dag}d_{o\dn}^{\dag}\bigr)|\hat{0}\rangle$
 &$\bar{U}\!-\!4(x^2\!+\!x'^2)\!-\!\tilde{\Delta}$
 &$U\!-\!16x^2\!-\!\frac{1}{6}y$ \\
&$2$ &$\frac{1}{\sqrt{2}}\bigl(d_{e\up}^{\dag}d_{o\dn}^{\dag}
   \!\!+\!d_{e\dn}^{\dag}d_{o\up}^{\dag}\bigr)|\hat{0}\rangle$
 &$U'\!-\!4x^2$ &$U\!-\!4x^2\!-\!\frac{1}{6}y$ \\
&$3$ &$\frac{1}{\sqrt{2}}\bigl(d_{e\up}^{\dag}d_{o\dn}^{\dag}
   \!\!-\!d_{e\dn}^{\dag}d_{o\up}^{\dag}\bigr)|\hat{0}\rangle$
 &$U\!-\!4x^2$ &$U\!-\!4x^2\!-\!\frac{1}{6}y$ \\
&$4$ &$d_{e\up}^{\dag}d_{o\up}^{\dag}|\hat{0}\rangle$
 &$U'\!-\!4x^2$ & $U\!-\!4x^2$ \\
&$5$ &$d_{e\dn}^{\dag}d_{o\dn}^{\dag}|\hat{0}\rangle$
 &$U'\!-\!4x^2$ & $U\!-\!4x^2$ \\
&$6$ &$\bigl(c_2 d_{e\up}^{\dag}d_{e\dn}^{\dag}
   \!\!-\!c_1 d_{o\up}^{\dag}d_{o\dn}^{\dag}\bigr)|\hat{0}\rangle$
 &$\bar{U}\!-\!4(x^2\!+\!x'^2)\!+\!\tilde{\Delta}$ &$U\!+\!\half y$
\end{tabular}
\end{center}
\caption{\label{molecular energies}
Low-lying eigenstates of $\hat{P}_0\hat{H}_{\el}\hat{P}_0$, where
$\hat{H}_{\el}$ describing the isolated molecule is defined in Eq.\
\eqref{hat H_el} and $\hat{P}_0$ is a projection operator into the sector of
the Fock space having occupancy $\hat{n}_b=0$ for the transformed phonon mode
defined in Eq.\ \eqref{hat b}.
Eigenstates \mbox{$|\phi^{(n_{\mol})}_i(\delta\!=\!=0)\rangle$} for
$\delta=\e_{\beta}=-\e_{\alpha}=0$ are grouped according to their total electron
number $n_{\mol}$, and specified in terms of operators $d_{p\sigma}^{\dag}$
defined in Eqs.\ \eqref{d_p} acting on $|\hat{0}\rangle$, the state having
$n_{\mol}=\hat{n}_b=0$;
$c_1$ and $c_2$ are real coefficients satisfying $c_1^2+c_2^2=1$ that reduce
for $U'=U$ to $c_1=1$, $c_2=0$. \mbox{$E^{(n_{\mol})}_i(\delta\!=\!0)$} is the
energy of state \mbox{$|\phi^{(n_{\mol})}_i(\delta\!=\!0)\rangle$}, expressed
in terms of $x=\lambda/\sqrt{\omega_0}$,
$x'=\lambda'/\sqrt{\omega_0}$, $\bar{U}=(U+U')/2$,
and $\tilde{\Delta}$ defined in Eq.\ \eqref{tilde Delta}.
\mbox{$E^{(n_{\mol})}_i(\delta\!>\!0)$} is the energy of the same state in the
special case $U'=U$ and $\lambda'=\lambda>0$, but including the leading
perturbative correction for $\delta>0$, expressed in terms of
$y=\omega_0(\delta/\lambda)^2 \exp[-4(\lambda/\omega_0)^2]$. For $U'=U$ and
$-\lambda'=\lambda>0$, the values \mbox{$E^{(n_{\mol})}_i(\delta\!>\!0)$} would
be the same apart from the interchange of the energies of the even- and
odd-parity $n_{\mol}=1$ states.}
\end{table}

Just as in the case of large $\e_{\beta}$, we begin by examining the low-lying
states of the isolated molecule, this time using the transformed Hamiltonian
$\hat{H}$ defined in Eq.\ \eqref{hat H} to find the energies. For the case
$U_{\alpha}=U_{\beta}=U$ considered throughout this section, $W=0$ in Eq.\
\eqref{hat H_el}. Then the only explicit e-ph coupling remaining in $\hat{H}$
enters through the terms
$t\sum_{\s} \bigl( B_{2\lambda'}^{\dag} d_{e\s}^{\dag} d_{o\s}^{\pdag} +
\mathrm{H.c.}\bigr)$ and $-J \bigl( B_{4\lambda'}^{\dag} I_e^+ I_o^- +
\mathrm{H.c.}\bigr)$. This subsection is concerned only with cases where
$|t|=\delta$ is small. If one also takes $|J|=\half|U'-U|$ to be small, then
the low-lying molecular states will contain only a weak admixture of components
having $\hat{n}_b>0$, where (as before) $\hat{n}_b$ is the number operator for
the transformed boson mode defined in Eq.\ \eqref{hat b}. Under this simplifying
assumption (which we re-examine in Sec.\ \ref{subsubsec:small-delta small-e_a}),
it suffices to focus on the eigenstates of $\hat{P}_0\hat{H}_{\el}\hat{P}_0$,
where $\hat{H}_{\el}$ given in Eq.\ \eqref{hat H_el} is the pure-electronic part
of $\hat{H}$, and $\hat{P}_0$ projects into the $\hat{n}_b=0$ Fock-space sector.
Table \ref{molecular energies} lists the low-lying energy eigenstates in this
sector for the case $\delta=0$ where the $\alpha$ and $\beta$ molecular
orbitals are exactly degenerate. Also listed are the energies of these states
for the special case $\lambda'=\lambda$ and $U'=U$, extended to include the
leading perturbative corrections for $\delta>0$. These corrections contain a
multiplicative factor $|\langle\hat{0}|B_{\pm 2\lambda'}|\hat{0}\rangle|^2 =
\exp[-4(\lambda/\omega_0)^2]$ (for $\lambda'=\lambda$) reflecting the reduction
with increasing e-ph coupling of the overlap of the phonon ground states for
Fock-space sectors of different $n_{\mol}$. Here and below, we denote by
$|\hat{0}\rangle$ the state having $n_{\mol}=\hat{n}_b=0$, which must be
distinguished from the state $|0\rangle$ in which $n_{\mol}=n_b=0$.

It can be seen from Table \ref{molecular energies} that for $\delta=0$ the
singly occupied sector has two states---depending on the sign of $\lambda'$,
either $|\phi^{(1)}_1\rangle$ and $|\phi^{(1)}_2\rangle$ or
$|\phi^{(1)}_3\rangle$ and $|\phi^{(1)}_4\rangle$---with lowest energy energy
$E^{(1)}_{\min}=-(\lambda+|\lambda'|)^2/\omega_0$. In cases of
small $|U'-U|$ and/or large $|\lambda'|$, the lowest state in the doubly
occupied sector is $|\phi_1^{(2)}\rangle$ with energy
$E^{(2)}_{\min}=\half(U+U')-4(\lambda^2+\lambda'^2)/\omega_0
-\tilde{\Delta}$, where
\begin{equation}
\label{tilde Delta}
\tilde{\Delta}=\sqrt{(8\lambda\lambda'/\omega_0)^2 + \tilde{J}^{\,2}} \, ,
\end{equation}
with
\begin{equation}
\label{tilde J}
\tilde{J} = J\, \bigl|\langle\hat{0}|B_{\pm 4\lambda'}|\hat{0}\rangle\bigr|^2
  = \half(U'-U)\,\exp(-8\lambda'^2/\omega_0^2) \, .
\end{equation}
One can use energies $E^{(1)}_{\min}$ and $E^{(2)}_{\min}$ to define an
effective Coulomb interaction
\begin{equation}
\label{tilde tilde U}
\tilde{\tilde{U}}=E^{(2)}_{\min}-2E^{(1)}_{\min} =
  \half(U+U')-\frac{2(\lambda-|\lambda'|)^2}{\omega_0} - \tilde{\Delta} .
\end{equation}
For $U'=U$, this value simplifies to
$\tilde{\tilde{U}}=U-2(\lambda+|\lambda'|)^2/\omega_0$, which decreases with
e-ph coupling at a greater rate than the effective Coulomb interaction
$\tilde{\tilde{U}}_{\alpha}$ [Eq.\ \eqref{tilde tilde U_a}] acting in the
$\alpha$ orbital when $\e_{\beta}-\e_{\alpha}$ is large.
The enhancement of e-ph renormalization of the Coulomb interaction in molecules
having multiple, nearly-degenerate orbitals improves the prospects of attaining
a regime of effective e-e attraction and may have interesting consequences in
the area of superconductivity.

Table \ref{molecular energies} also indicates that the ground state of the
isolated molecule crosses from single electron occupancy (for weaker e-ph
couplings) to double occupancy (for stronger e-ph couplings) at the point where
$E^{(2)}_{\min}=E^{(1)}_{\min}$, which reduces for $\delta=0$
and small $\tilde{J}$ to
\begin{equation}
\label{lambda_x}
\frac{(\lambda+|\lambda'|)^2}{\omega_0} = \frac{U+U'}{6} .
\end{equation}
Just as in cases where the $\beta$ molecular orbital lies far above the Fermi
energy of the leads (Sec. \ref{subsec:large-e_b}), we will see that this level
crossing in the isolated molecule is closely connected to a crossover in the
full problem that results in pronounced changes in the system's low-temperature
properties. The lowest energy of any molecular state having three electrons
(not shown in Table \ref{molecular energies}) is
$E^{(3)}_{\min}(\delta\!=\!0)=U+2U'-(3\lambda+|\lambda'|)^2/\omega_0$,
while the sole four-electron state has energy
$E^{(4)}_1(\delta\!=\!0)=2U+4U'-16\lambda^2/\omega_0$.
For all the cases considered in Figs.\
\ref{small-delta props}--\ref{occs vs gate} below, these energies are
sufficiently high that states with $n_{\mol}>2$ play no role in the
low-energy physics.

\begin{figure}[tb]
\centering\includegraphics[width=3.25in]{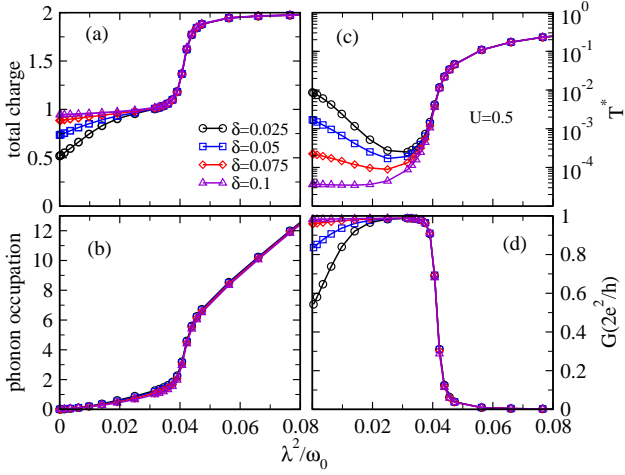}
\caption{\label{small-delta props} (Color online)
Variation with $\lambda^2/\omega_0$ of (a) the ground-state molecular charge
$\langle n_{\mol}\rangle=\langle n_e+n_o\rangle$, (b) the ground-state phonon
occupation $\langle n_b\rangle$, (c) the crossover temperature $T^*$, and (d)
the zero-temperature linear conductance $G$, all calculated for $U'=U=0.5$,
$\lambda'=\lambda$, and the four values of $\delta=\e_{\beta}=-\e_{\alpha}$
listed in the legend. In the case $\delta=0.05$, the orbital energy splitting
is in resonance with the phonon energy, i.e.,
$\e_{\beta}-\e_{\alpha}=2\delta=\omega_0$.}
\end{figure}

\begin{figure}[tb]
\centering\includegraphics[width=3.5in]{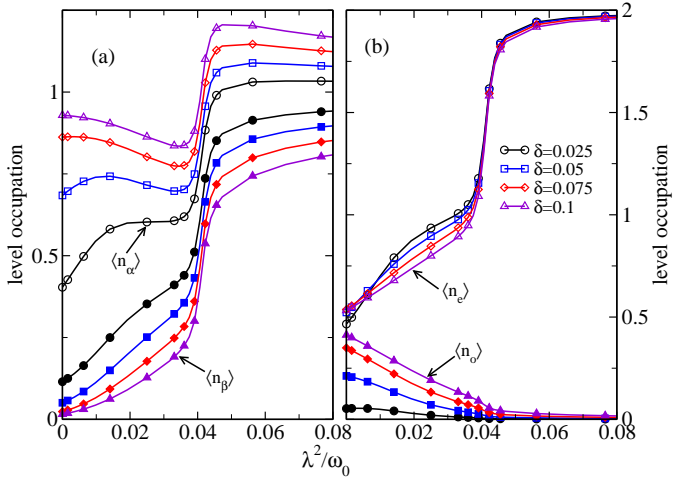}
\caption{\label{small-delta charges} (Color online)
Occupation of individual molecular orbitals vs $\lambda^2/\omega_0$ for
$U'=U=0.5$, $\lambda'=\lambda$, and the four values of
$\delta=\e_{\beta}=-\e_{\alpha}$ listed in the legend:
(a) $\langle n_\alpha\rangle$ (open symbols) and $\langle n_\beta\rangle$
(filled symbols); (b) $\langle n_e\rangle$ (open symbols) and
$\langle n_o\rangle$ (filled symbols).}
\end{figure}

\subsubsection{Both orbitals close to the chemical potential}
\label{subsubsec:small-delta small-e_a}

This subsection presents numerical solutions of the full problem under variation
of the e-ph coupling. As before, we focus primarily on the reference case
$\lambda'=\lambda$, $U'=U=0.5$.

Figure \ref{small-delta props} plots the evolution with $\lambda^2/\omega_0$
of the same properties as appear in Fig.\ \ref{large-e_b props} for four values
of $\delta$ chosen so that the two figures differ only as to the energy of the
upper molecular orbital: $\e_{\beta}=4\gg U$ in the earlier figure versus
$\e_{\beta}=\delta\ll U$ here. The results in the two figures are superficially
similar, although there are some significant differences as will be explained
below.

We begin by considering the behavior for $\lambda=0$. Figure
\ref{small-delta props}(a) shows the zero-temperature molecular charge
$\langle n_{\mol}\rangle$, while Fig.\ \ref{small-delta charges} displays
the corresponding occupancies of individual molecular orbitals:
$\langle n_{\alpha}\rangle$ and $\langle n_{\beta}\rangle$ in panel (a), and
$\langle n_e\rangle$ and $\langle n_o\rangle$ in panel (b). For $\delta\ll
\Gamma=\pi V^2/D\simeq 0.0177$, $\langle n_{\mol}\rangle\simeq\langle n_e\rangle
\simeq 0.5$, which may be understood as a consequence of the ground state being
close to that for $U=V=\infty$ and $\delta=0$: a product of (1)
$\half \bigl[ c_{e\up}^{\dag} d_{e\dn}^{\dag} - c_{e\dn}^{\dag} d_{e\up}^{\dag}
- \sqrt{2} c_{e\up}^{\dag} c_{e\dn}^{\dag} \bigr] |0\rangle$
where $c_{e\s}=(2N_s)^{-1/2}\sum_{\bk}c_{e\bk\s}$ annihilates an electron
in the linear combination of left- and right-lead states that tunnels into/out
of the molecular orbitals, and (2) other lead degrees of freedom that are
decoupled from the molecule. The total charge increases with $\delta$
and approaches $\langle n_{\mol}\rangle=\langle n_{\alpha}\rangle=1$ for
$\delta\gg\Gamma$, in which limit the large Coulomb repulsion $U$ leads to
local-moment formation in the $\alpha$ orbital, followed at low temperatures
by Kondo screening, very much in the same manner as found for $\e_{\beta}=4$
(Sec.\ \ref{subsubsec:large-e_b small-e_a}).

Turning on e-ph couplings $\lambda'=\lambda$ lowers the energy of the
even-parity molecular orbital below that of the odd orbital, and initially
drives the system toward $\langle n_e\rangle=1$, $\langle n_o\rangle=0$, and
toward a many-body singlet ground state formed between the leads and a local
moment in the even-parity molecular orbital (rather than the local moment
in the $\alpha$ orbital that is found for $\e_{\beta}=4$). The spin-screening
scale $T^*$ in Fig.\ \ref{small-delta props}(c) shows an initial decrease with
increasing $\lambda^2/\omega_0$ that is very strong for the smaller values for
$\delta$, where the e-ph coupling drives the system from mixed valence into
the Kondo regime. For larger $\delta$, where the system is in the Kondo limit
even at $\lambda=0$, there is a much milder reduction of $T^*$ caused by the
phonon-induced shift of the filled molecular orbital further below the chemical
potential.

Upon further increase in the e-ph coupling, $\langle n_{\mol}\rangle$ and $T^*$
both show rapid but continuous rises around some value $\lambda=\lambda_x$.
The crossover value $\lambda_x^2/\omega_0\simeq 0.042$, which is independent of
$\delta$ for $\delta\ll U$, coincides closely with its $\delta=0$ value
$U/12\simeq 0.0417$ for the isolated molecule, where it describes the crossing
of the singly occupied state $|\phi_1^{(1)}\rangle$ and the doubly occupied
state $|\phi_1^{(2)}\rangle$ (see Table \ref{molecular energies}). For
$\Gamma>0$, the crossover of the ground-state molecular charge from 1 to 2 is
smeared over the range $|U-12\lambda^2/\omega_0|\lesssim 2\Gamma$, suggesting
a full width for the crossover $\Delta(\lambda^2/\omega_0)\simeq 4\Gamma/12 =
0.006$, in good agreement with Figs.\ \ref{small-delta props}(a) and
\ref{small-delta charges}. The values of $\lambda_x$ and
$\Delta(\lambda^2/\omega_0)$ are smaller than the corresponding values for
$\e_{\beta}=4$ by factors of roughly 3 and 2, respectively, a consequence of the
stronger e-ph effects found for small molecular orbital energy separation.
Moreover, the absence of any dependence of $\lambda_x$ on $\delta$ is to be
contrasted with the linear dependence of the crossover e-ph coupling on
$\e_{\alpha}$ in Fig.\ \ref{large-e_b props}.

In the regime $\lambda\gtrsim\lambda_x$, the system minimizes the e-ph
energy by adopting orbital occupancies $\langle n_e\rangle\simeq 2$,
$\langle n_o\rangle\simeq 0$ (shown in Fig.\ \ref{small-delta charges} to hold
for all the $\delta$ values considered). Here, $T^*$ approaches the scale
$12\lambda^2/\omega_0-U$ at which occupation of $n_{\mol}=1$ molecular states
becomes frozen out.
Over the entire range of $\delta$ and $\lambda^2/\omega_0$ illustrated in Figs.\
\ref{small-delta props} and \ref{small-delta charges}, the ground-state phonon
occupation $\langle n_b\rangle$ [Fig.\ \ref{small-delta props}(b)] closely
tracks $\bar{n}_b$ and the $T=0$ conductance [Fig.\ \ref{small-delta props}(d)] is
everywhere well-described by Eq.\ \eqref{FL cond}.

We note that the equilibrium properties shown in Figs.\ \ref{small-delta props}
and \ref{small-delta charges} exhibit no special features in the resonant case
$\delta=0.05$ in which the molecular orbital spacing $\e_{\beta}-\e_{\alpha}$
exactly matches the phonon energy. We expect the resonance condition to play
a significant role only in driven setups where a nonequilibrium phonon
distribution serves as a net source or sink of energy for the electron
subsystem.

\begin{figure}[tb]
\centering\includegraphics[width=3.3in]{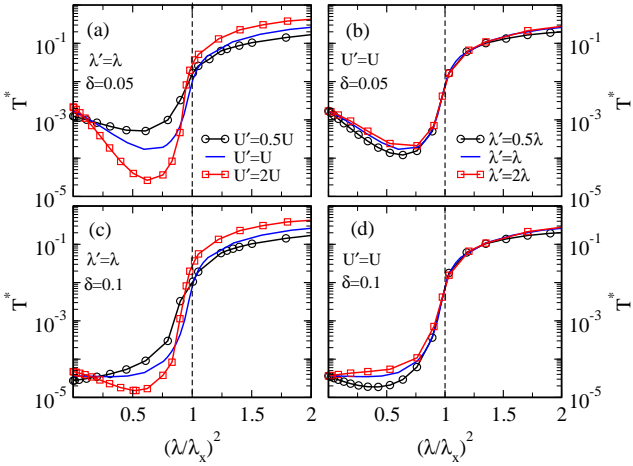}
\caption{\label{T* vs scaled lambda} (Color online)
Crossover temperature $T^*$ vs scaled e-ph coupling $(\lambda/\lambda_x)^2$.
The left panels show different ratios $U'/U$ for $\lambda'=\lambda$ while
the right panels show different $\lambda'/\lambda$ for $U'=U$. The upper panels
(a), (b) correspond to an orbital separation $\delta=0.05$, and the lower
panels (c), (d) treat $\delta=0.1$. All data are for $U=0.5$. Vertical dashed
lines at $\lambda=\lambda_x$ [calculated via the condition
$E^{(2)}_{\min}(\delta=0)=E^{(1)}_{\min}(\delta=0)$] separate the
Kondo regime from the phonon-dominated regime.}
\end{figure}

The properties presented above are little changed under relaxation of the
assumptions $\lambda'=\lambda$ and $U'=U$. For reasons of space, we show data
only for the variation of the crossover temperature $T^*$ with e-ph coupling
with different fixed values of $U'/U$ [Figs.\ \ref{T* vs scaled lambda}(a) and
\ref{T* vs scaled lambda}(c)] or $\lambda'/\lambda$ [Figs.\
\ref{T* vs scaled lambda}(b) and \ref{T* vs scaled lambda}(d)]. In each case,
$T^*$ is plotted against $(\lambda/\lambda_x)^2$, where $\lambda_x$ is the
value of $\lambda$ that satisfies the condition
$E^{(2)}_{\min}(\delta=0)=E^{(1)}_{\min}(\delta=0)$ for crossover from single
to double occupation of the isolated molecule. For $U'=0.5U$ and $U'=2U$, it
must be recognized that $J=\half(U'-U)$ is not small, calling into question the
validity of the approximation $\hat{n}_b=0$ used to derive the energies in
Table \ref{molecular energies}. What is more, the data shown are for nonzero
orbital energy splittings $\delta=0.05$ (upper panels) and $\delta=0.1$
(lower panels). Nonetheless, the plots all exhibit good data collapse along the
horizontal axis, showing that $\lambda_x$ calculated for $\hat{n}_b=0$
and $\delta=0$ captures very well the scale characterizing the crossover from
the Kondo regime ($\lambda\lesssim\lambda_x$) to the phonon-dominated regime
($\lambda\gtrsim\lambda_x$).

The data in Fig.\ \ref{T* vs scaled lambda} show greater spread along the
vertical axis, particularly in the Kondo regime under variation of $U'/U$.
However, we find that in each panel, the value of $T^*$ in the phonon-dominated
regime can be reproduced with good quantitative accuracy by applying the
condition $T^*\chi(T*)=0.0701$ to the susceptibility of the isolated molecule,
calculated using the eleven states listed in Table \ref{molecular energies}.
This provides further evidence for the adequacy of the approximation
$\hat{n}_b=0$ employed in the construction of the table. More importantly,
Fig.\ \ref{T* vs scaled lambda} shows that the physics probed in Figs.\
\ref{small-delta props} and \ref{small-delta charges} for the special case
$\lambda'=\lambda$ and $U'=U$ is broadly representative of the behavior
over a wide region of the model's parameter space.

\begin{figure}[tb]
\centering\includegraphics[width=3.25in]{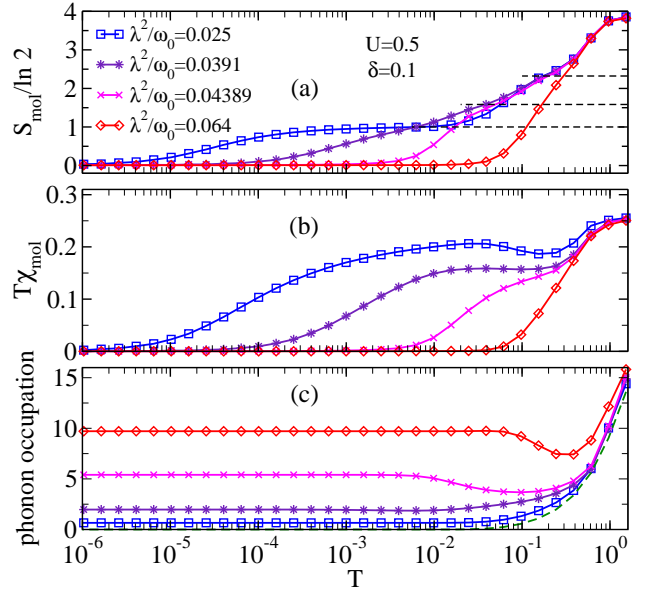}
\caption{\label{small-e_b thermo} (Color online)
Temperature dependence of (a) the molecular entropy, (b) temperature times the
molecular susceptibility, $T\chi_{\mol}\equiv|\bm{\mu}_{\mol}|^2/3$, where
$\bm{\mu}_{\mol}$ is the molecule's magnetic moment, and (c) the phonon
occupation. Data are for $U'=U=0.5$, $\delta=0.1$, $\lambda'=\lambda$, and the
four values of $\lambda^2/\omega_0$ listed in the legend. In (a), the horizontal
dashed lines mark $S_{\mol}=\ln2$, $\ln 3$, and $\ln 5$. In (c), the dashed
line shows the occupation of a free phonon mode of energy $\omega_0=0.1$.}
\end{figure}

To this point, we have concentrated on ground-state ($T=0$) properties and the
temperature scale $T^*$ characterizing the quenching of the molecular magnetic
moment. We now illustrate the full temperature dependence of three thermodynamic
properties in situations where the molecular orbitals are arranged symmetrically
around the chemical potential. Figure \ref{small-e_b thermo} plots the variation
with $T$ of the molecular entropy, molecular susceptibility, and phonon
occupation for $U'=U=0.5$, $\delta=0.1$, $V=0.075$, and four different values of
$\lambda'=\lambda$.
As long as the temperature exceeds all molecular energy scales, the entropy and
susceptibility are close to the values $S_{\mol}=\ln 4$ and
$T\chi_{\mol}=1/8$ attained when every one of the 16 molecular
configurations has equal occupation probability, while the phonon occupation is
close to the Bose-Einstein result for a free boson mode of energy $\omega_0$
[dashed line in Fig.\ \ref{small-e_b thermo}(c)]. Once the temperature drops
below $U$, most of the molecular configurations (and all with total charge
$n_{\mol}>2$) become frozen out. For $\lambda\ll\lambda_x$ (exemplified by
$\lambda^2/\omega_0=0.025$ in Fig. \ref{small-e_b thermo}), there is a slight
shoulder in the entropy around $S_{\mol}=\ln 5$ and a minimum in the square of
the local moment around $T\chi_{\mol}=1/5$, the values expected when the empty
and singly occupied molecular configurations (the first five states listed in
Table \ref{molecular energies} are quasidegenerate. At lower temperatures,
there is an extended range of local-moment behavior ($S_{\mol}=\ln 2$,
$T\chi_{\mol}\simeq 1/4$) associated with single occupancy of the even-parity
molecular orbital (states $|\phi^{(1)}_1\rangle$ and $|\phi^{(1)}_2\rangle$).
Eventually, the properties cross over below the temperature scale $T^*$ defined
above to those of the Kondo singlet state: $S_{\mol}=0$, $T\chi_{\mol}=0$.

\begin{figure}[tb]
\centering\includegraphics[width=3.25in]{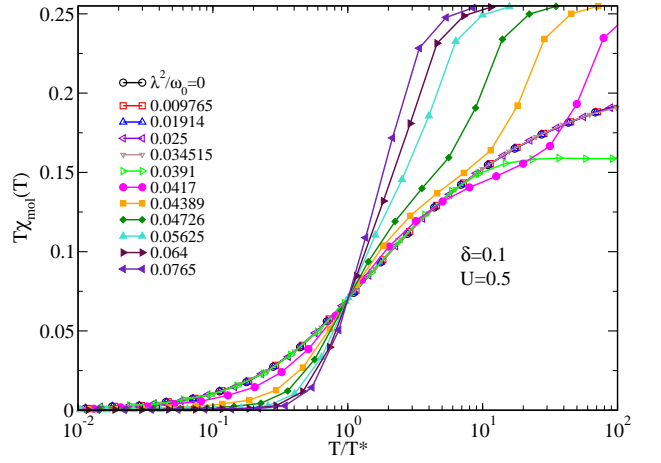}
\caption{\label{universality} (Color online)
Magnetic moment $\mu^2=T\chi_{\mol}$ vs scaled temperature $T/T^*$ for
$U'=U=0.5$, $\lambda'=\lambda$, $\delta=0.1$, and values of
$\lambda^2/\omega_0$ spanning the crossover from the Kondo regime to the doubly
occupied regime. The collapse over the range $T\lesssim 10T^*$ of all curves
corresponding to $\lambda^2/\omega_0 \le 0.0391$ demonstrates the universal
physics of the Kondo regime. No such universality is present in the
boson-dominated limit.}
\end{figure}

For $\lambda$ just below $\lambda_x$ ($\lambda^2/\omega_0 = 0.0391$ in Fig.\
\ref{small-e_b thermo}) there are weak shoulders near $S_{\mol}=\ln 5$ and
$T\chi_{\mol}=1/5$, as in the limit of smaller e-ph couplings. In this case,
however, these features reflect the near degeneracy of the four $n_{\mol}=1$
configurations and the lowest-energy $n_{\mol}=2$ configuration:
$|\phi^{(2)}_1\rangle$ in Table \ref{molecular energies}. At slightly lower
temperatures, the states $|\phi^{(1)}_3\rangle$ and $|\phi^{(1)}_4\rangle$
become depopulated and the properties drop through $S_{\mol}=\ln 3$ and
$T\chi_{\mol}=1/6$ before finally falling smoothly to zero. Even though there
is no extended regime of local-moment behavior, the asymptotic approach of
$S_{\mol}$ and $T\chi_{\mol}$ to their ground state values is essentially
identical to that for $\lambda\ll\lambda_x$ after rescaling of the temperature
by $T^*$. As shown in Fig.\ \ref{universality}, throughout the regime
$\lambda<\lambda_x$, $T\chi_{\mol}$ follows the same function of $T/T^*$ for
$T\lesssim 10T^*$. This is just one manifestation of the universality of the
Kondo regime, in which $T_K\equiv T^*$ serves as the sole low-energy scale.

A small increase in $\lambda^2/\omega_0$ from $0.0391$ to $0.04389$, slightly
above $\lambda_x^2/\omega_0=0.0417$, brings about significant changes in the
low-temperature properties. While there are still weak features in the entropy
at $\ln 5$ and $\ln$3, the final approach to the ground state is more rapid
than for $\lambda<\lambda_x$, as can be seen from Fig.\ \ref{universality}.
Note also the upturn in $\langle n_b\rangle$ as $T$ falls below about
$10T^*$---a feature absent for $\lambda <\lambda_x$ that signals the integral
role played by phonons in quenching the molecular magnetic moment.

Finally, in the limit $\lambda\gg\lambda_x$ (exemplified by
$\lambda^2/\omega_0=0.064$ in Fig.\ \ref{small-e_b thermo}), $E^{(2)}_1$ is by
a considerable margin the lowest eigenvalue of
$\hat{P}_0\hat{H}_{\el}\hat{P}_0$, so with decreasing temperature, $S_{\mol}$
and $T\chi_{\mol}$ quickly approach zero with little sign of any intermediate
regime. Even though the quenching of the molecular degrees of freedom arises
from phonon-induced shifts in the molecular orbitals rather than from a
many-body Kondo effect involving strong entanglement with the lead degrees, the
$\lambda\to\infty$ ground state is adiabatically connected to that for
$\lambda=0$.

\begin{figure}[tb]
\centering\includegraphics[width=3.3in]{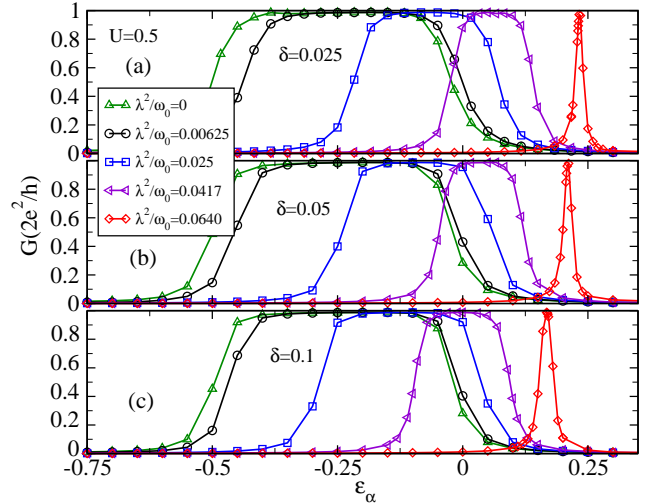}
\caption{ (Color online)\label{cond vs gate}
Zero-temperature conductance $G$ as a function of $\e_{\alpha}=-\delta-V_g$
(where $V_g$ is an applied gate voltage) for the five values of
$\lambda^2/\omega_0$ listed in the legend and (a) $\delta=0.025$,
(b) $\delta=0.05$, and (c) $\delta=0.1$.
The other parameters are $U'=U=0.5$ and $\lambda'=\lambda$.}
\end{figure}

\begin{figure}[tb]
\centering\includegraphics[width=3.3in]{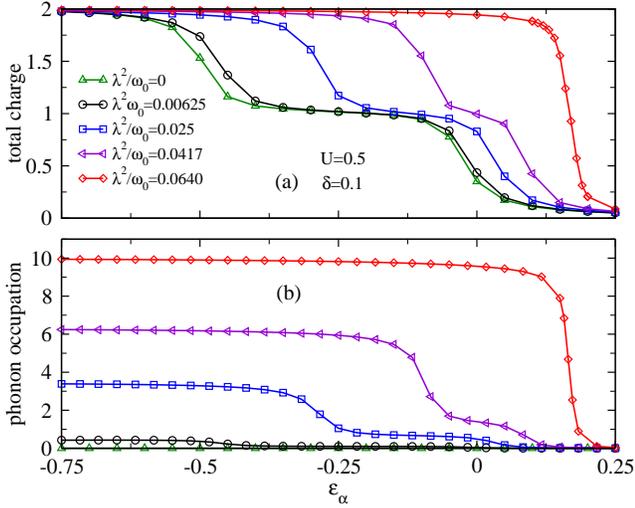}
\caption{\label{occs vs gate} (Color online)
(a) Ground-state molecular charge and (b) ground-state phonon occupation
as functions of $\e_{\alpha}=-\delta-V_g$ (where $V_g$ is an applied gate
voltage) for $\delta=0.1$ and the values of $\lambda^2/\omega_0$ listed in the
legend. The other parameters are $U'=U=0.5$ and $\lambda'=\lambda$.}
\end{figure}

\subsubsection{Effect of a uniform shift in the orbital energies}
\label{subsubsec:small-delta sweeps}

We finish by considering the effect of shifting the two molecular orbitals at
a fixed, small energy separation $\e_{\beta}-\e_{\alpha}=2\delta$ through the
application of a gate voltage $V_g$ that causes $\e_i$ in Eq.\ \eqref{H_el}
to be replaced by $\e_i-eV_g$, and $\tilde{\e}_p$ in Eq.\ \eqref{hat H_el}
to be replaced by $\tilde{\e}_p-eV_g$.
Figure \ref{cond vs gate} plots the gate-voltage dependence of the linear
conductance for $U'=U=0.5$, five values of $\lambda'=\lambda$, and for
$\delta=0.025$ [panel (a)], $\delta=0.05$ (b), and $\delta=0.1$ (c).
Figure \ref{occs vs gate} shows the corresponding evolution of the total
molecular charge and the phonon occupation for the case $\delta=0.1$. In both
figures, the quantity plotted along the horizontal axis is
$\e_{\alpha}=-\delta-eV_g$, which allows direct comparison with the results
shown in in Figs.\ \ref{large-e_b occs}(c), \ref{large-e_b occs}(f), and
\ref{large-e_b cond}(c) for the regime where the $\beta$ molecular orbital lies
far above the chemical potential.

Just as in the other situations considered above, the zero-temperature
conductance obeys the Fermi-liquid relation Eq. \eqref{FL cond}. A plateau at
$G\simeq G_0$ spans the range of gate voltages within which the total molecular
occupancy is $\langle n_{\mol}\rangle\simeq 1$ [e.g., compare Figs.\
\ref{cond vs gate}(c) and \ref{occs vs gate}(a)], while the conductance
approaches zero for larger $V_g$, where the molecular charge vanishes, and for
smaller $V_g$, where $\langle n_{\mol}\rangle\simeq 2$.

Once again, we begin by considering the limit $\lambda=0$ of zero e-ph coupling.
For $\delta=0.025\ll\Gamma=0.0177$, the rises between zero and peak conductance
are somewhat broader (along the $\e_{\alpha}$ axis) than their counterparts in
cases where the $\beta$ molecular orbital lies far above the chemical potential
[compare with Fig. \ref{large-e_b cond}(c)]. This broadening can be understood
as a consequence of the step in $\langle n_{\mol}\rangle$ being split into
changes in $\langle n_{\alpha}\rangle$ and in $\langle n_{\beta}\rangle$. When
$\delta\gg\Gamma$, the $\beta$ molecular orbital is essentially depopulated
[as can be seen for the $V_g=0$ in Fig.\ \ref{small-delta charges}(a)] and
the conductance steps narrow to a width similar to that for $\e_{\beta}=4$.

Increase of the e-ph coupling from zero results in shifts of the occupancy and
conductance steps to progressively higher values of $\e_{\alpha}$ (or to lower
values of $V_g$) that can be attributed to the phonon-induce renormalization
of the orbital energies and of the Coulomb interactions. For $\delta=0.025$,
the width of the $\langle n_{\mol}\rangle\simeq 1$, $G\simeq G_0$ plateau is
close to the value $\tilde{\tilde{U}}$ defined in Eq.\ \ref{tilde tilde U},
which approaches $U-8\lambda^2/\omega_0$ in the limit
$\delta\lesssim\lambda^2/\omega_0$ satisfied by the $\delta=0.025$ curves in
Eq.\ \ref{cond vs gate}(a). Even for the $\delta=0.1$ curves shown in
Fig.\ \ref{cond vs gate}(b), the plateau width is at least
$U-8\lambda^2/\omega_0$, considerably larger than than its value
$\tilde{U}=U-2\lambda^2/\omega_0$ when the orbital $\beta$ lies far above the
chemical potential. The occupancy and conductance plateau might be expected to
disappear once $\tilde{\tilde{U}}$ becomes negative around
$\lambda^2/\omega_0\simeq U/8=0.0625$. Indeed, the data for
$\lambda^2/\omega_0=0.064$ in Fig.\ \ref{cond vs gate} show a narrow
conductance peak that can be associated with the rapid decrease of
$\langle n_{\mol}\rangle$ directly from 2 to 0 without any significant range of
single occupancy [illustrated for $\delta=0.1$ in Fig.\ \ref{occs vs gate}(b)].

\section{Summary}
\label{sec:summary}

We have studied the low-temperature properties of a single-molecule junction
formed by a two-orbital molecule connecting metallic leads. The model
Hamiltonian incorporates intraorbital and interorbital Coulomb repulsion, a
Holstein coupling of the molecular charge to the displacement of a local
phonon mode, and also phonon-mediated interorbital tunneling. We have
investigated the low-temperature regime of the system using the numerical
renormalization group to provide a nonperturbative treatment of the competing
strong interactions. Insight into the numerical results has been obtained by
considering the phonon-renormalization of model parameters identified through
canonical transformation of the starting Hamiltonian.

We have focused on two quite different regions of the model's parameter space:
(1) In situations where one of the two molecular orbitals lies close to the
chemical potential while the other has a much higher energy, the thermodynamic
properties and linear conductance are very similar to those predicted
previously for a single-orbital molecule, showing phonon-induced shifts in the
active molecular orbital and a reduction in the effective Coulomb repulsion
between electrons on the molecule. In this region, interorbital e-ph
coupling can be treated as a weak perturbation. (2) In the region in which the
two orbitals both lie close to the chemical potential, where all the
interactions must be treated on an equal footing, the phonon-induced
renormalization of the Coulomb interactions is stronger than in the case of one
active molecular orbital, enhancing the likelihood of attaining in experiments
the interesting regime of small or even attractive on-site Coulomb interactions.

In both regions (1) and (2), electron-phonon interactions favor double
occupancy of the molecule and are detrimental to formation of a molecular
local moment and to the low-temperature Kondo screening of that moment by
electrons in the leads. With increasing electron-phonon coupling, the Kondo
effect is progressively destroyed and a phonon-dominated nonmagnetic ground
state emerges in its place. In all the cases presented here, this evolution
produces a smooth crossover in the ground-state properties.
Special situations that result in first-order quantum phase transitions between
Kondo and non-Kondo ground states will be described in a subsequent publication.
We have left for future study cases involving two degenerate (or nearly
degenerate) molecular orbitals lying below the chemical potential of the leads.
In such cases, e-e interactions favor the presence of an unpaired electron in
each orbital, and electron-phonon interactions may be expected to significantly
affect the competition between total-spin-singlet and triplet ground
states.\cite{Hofstetter:2002,Gordon:2003,Nature:453-366}

\acknowledgments

The authors acknowledge partial support of this work by CAPES (G.I.L), by CNPq
under grant 493299/2010-3 (E.V) and a CIAM grant (E.V.\ and E.V.A.), by
FAPEMIG under grant CEX-APQ-02371-10 (E.V), FAPERJ (E.V.A), and by the NSF
Materials World Network program under grants DMR-0710540 and DMR-1107814 (L.D.\
and K.I.).

\end{document}